\documentclass{IEEEtran}

\usepackage{soul}

\usepackage{stmaryrd}
\usepackage{amsmath}
\usepackage{amsthm}
\usepackage{amssymb}
\usepackage{graphicx}
\usepackage{todonotes}
\usepackage{cite}
\usepackage{algorithmic}
\usepackage{algorithm}
\usepackage{mathtools}
\usepackage{caption}
\usepackage{subcaption}
\usepackage{float}
\usepackage{comment}
\usepackage{makecell}
\usepackage{xcolor}
\usepackage{multirow}
\usepackage{bm}
\usepackage{bbm}
\usepackage{glossaries}
\usepackage{enumitem}
\usepackage{extarrows}
\usepackage{tikz}
\usetikzlibrary{arrows.meta, positioning, shapes.geometric}
\tikzset{
	block/.style={draw, thick, minimum height=0.8cm, minimum width=1.2cm, align=center},
    amp/.style={draw, thick, shape=isosceles triangle, isosceles triangle apex angle=60, minimum height=0.7cm, anchor=apex},
    circ/.style={draw, circle, inner sep=1pt, fill=black},
    antenna/.pic={
    \draw[thick] 
      (0,0) -- ++(0,0.6) -- ++(-0.35,0.4)
             (0,0) -- ++(0,0.6) -- ++(0,0.4)
             (0,0) -- ++(0,0.6) -- ++(0.35,0.4) -- ++(-0.7,0);

  	}
}

\newcommand{\vect}[1]{\boldsymbol{#1}}
\newcommand{\mat}[1]{\boldsymbol{#1}}

\newcommand{\adj}{\mathrm{adj}}

\newcommand{\norm}[1]{\left\lVert#1\right\rVert}
\newcommand{\matnorm}[1]{{\left\vert\kern-0.25ex\left\vert\kern-0.25ex\left\vert #1 \right\vert\kern-0.25ex\right\vert\kern-0.25ex\right\vert}}
\newcommand{\abs}[1]{\left\lvert#1\right\rvert}

\newcommand{\cn}[2]{\ensuremath{\mathcal{C}\cN\left(#1,#2\right)}}

\newcommand{\maximize}{\mathop{\rm maximize}\limits}
\newcommand{\subjectto}{\mathop{\rm s.t.}\limits}
\newcommand{\minimize}{\mathop{\rm minimize}\limits}

\newcommand{\prob}[1][]{
\ifthenelse{\isempty{#1}}%
      {\ensuremath{P}}%
    {\ensuremath{P\left\(#1\right\)}}%
}

\def\defeq{\triangleq}

\renewcommand{\Re}{\operatorname{Re}}

\newcommand{\transp}{{\sf T}}
\newcommand{\herm}{{\sf H}}
\newcommand{\sinr}{{\sf SINR}}
\newcommand{\opt}{\mathtt{opt}}
\newcommand{\mmse}{\mathtt{mmse}}

\def\ba{{\vect{a}}}

\def\bc{{\vect{c}}}

\def\bg{{\vect{g}}}
\def\bh{{\vect{h}}}

\def\bq{{\vect{q}}}
\def\br{{\vect{r}}}

\def\bu{{\vect{u}}}

\def\bw{{\vect{w}}}
\def\bx{{\vect{x}}}
\def\by{{\vect{y}}}

\def\b0{{\vect{0}}}

\def\bzero{{\vect{0}}}

\def\bphi{{\vect{\phi}}}
\def\bpsi{{\vect{\psi}}}

\def\brho{{\vect{\rho}}}
\def\bvarpi{\vect{\varpi}}
\def\balpha{\vect{\alpha}}

\def\bvarpi{\vect{\varpi}}

\def\bD{{\mat{D}}}

\def\bG{{\mat{G}}}
\def\bH{{\mat{H}}}
\def\bI{{\mat{I}}}

\def\bR{{\mat{R}}}

\def\bX{{\mat{X}}}
\def\bY{{\mat{Y}}}
\def\bZ{{\mat{Z}}}

\def\bPhi{{\mat{\Phi}}}
\def\bGamma{\mat{\Gamma}}

\def\bSigma{\mat{\Sigma}}

\def\cH{{\mathcal{H}}}

\def\cN{{\mathcal{N}}}

\def\cS{{\mathcal{S}}}

\def\E{{\mathbb{E}}}

\def\C{{\mathbb{C}}}
\def\R{{\mathbb{R}}}

\def\sfG{{\mathsf{G}}}

\def\sfZ{{\mathsf{Z}}}

\def\sfa{{\mathsf{a}}}

\def\sfh{{\mathsf{h}}}

\def\sfp{{\mathsf{p}}}

\def\sfr{{\mathsf{r}}}

\def\sfu{{\mathsf{u}}}

\def\sfx{{\mathsf{x}}}

\def\ttB{\mathtt{B}}

\def\ttD{\mathtt{D}}

\def\ttR{\mathtt{R}}

\def\ttU{\mathtt{U}}

\def\ul{\text{ul}}
\def\dl{\text{dl}}

\theoremstyle{plain}
\newtheorem{theorem}{Theorem}
\newtheorem{propi}{Proposition}
\newtheorem{corollary}{Corollary}
\newtheorem{lemma}{Lemma}

\theoremstyle{remark}
\newtheorem{remark}{Remark}

\theoremstyle{definition}
\newtheorem{assumption}{Assumption}
\newtheorem{definition}{Definition}
\newenvironment{proof outline}{\paragraph*{Proof Outline}}{\hfill$\IEEEQEDopen$}

\newcommand{\algorithmicinitialize}{\textbf{Initialize:}}
\newcommand{\INITIALIZE}{\item[\algorithmicinitialize]}

\newacronym{mimo}{MIMO}{multiple-intput multiple-output}
\newacronym{bs}{BS}{base station}
\newacronym{lti}{LTI}{linear time-invariant}
\newacronym{cir}{CIR}{channel impulse response}
\newacronym{cfr}{CFR}{channel frequency response}
\newacronym{iid}{i.i.d.}{independent and identically distributed}
\newacronym{svd}{SVD}{singular value decomposition}
\newacronym{cpu}{CPU}{central processing unit}
\newacronym{sinr}{SINR}{signal-to-interference-plus-noise ratio}
\newacronym{snr}{SNR}{signal-to-noise ratio}
\newacronym{los}{LoS}{line-of-sight}
\newacronym{lsfc}{LSFC}{large-scale fading coefficient}
\newacronym{bibo}{BIBO}{bounded-input bounded-output}
\newacronym{tdl}{TDL}{tapped delay line}
\newacronym{mmse}{MMSE}{minimum mean-square error}
\newacronym{zf}{ZF}{zero forcing}
\newacronym{mr}{MR}{maximum ratio}
\newacronym{sic}{SIC}{successive interference cancellation}
\newacronym{dft}{DFT}{discrete Fourier transform}
\newacronym{awgn}{AWGN}{additive white Gaussian noise}
\newacronym{tdd}{TDD}{time-division duplex}
\newacronym{ap}{AP}{access point}
\newacronym{ofdm}{OFDM}{orthogonal frequency-division multiplexing}
\newacronym{csi}{CSI}{channel state information}
\newacronym{mse}{MSE}{mean-square error}
\newacronym{psd}{PSD}{positive semi-definite}
\newacronym{qp}{QP}{quadratic program}
\newacronym{uma}{UMa}{urban macro}
\newacronym{umi}{UMi}{urban micro}
\newacronym{cdf}{CDF}{cumulative distribution function}
\newacronym{ncr}{NCR}{network-controlled repeater}
\newacronym{ris}{RIS}{reflective intelligent surface}
\newacronym{iab}{IAB}{integrated access and backhaul}
\newacronym{nr}{NR}{New Radio}
\newacronym{u2r}{U2R}{user-to-repeater}
\newacronym{r2b}{R2B}{repeater-to-BS}
\newacronym{r2r}{R2R}{inter-repeater}

\makeglossaries

\makeatletter
\renewcommand*\env@matrix[1][\arraystretch]{%
  \edef\arraystretch{#1}%
  \hskip -\arraycolsep
  \let\@ifnextchar\new@ifnextchar
  \array{*\c@MaxMatrixCols c}}
\makeatother

\usepackage{url}
\usepackage[hidelinks, breaklinks]{hyperref}
\hypersetup{
	colorlinks   = true, 
	urlcolor     = black, 
	linkcolor    = black, 
	citecolor   = black, 
}

\usepackage[capitalise,nameinlink]{cleveref}  
\usepackage{prettyref}
 
\crefformat{equation}{(#2#1#3)} 
\crefname{figure}{Fig.}{Figs.}
\crefname{section}{Sec.}{Secs.}

\usepackage[normalem]{ulem}

\begin{document}

\title{Repeater Swarm-Assisted Cellular Systems:\\ Interaction Stability and Performance Analysis
	
\author{Jianan Bai, Anubhab Chowdhury, Anders Hansson, and Erik G. Larsson}
\thanks{Parts of the results in this paper were presented at IEEE SPAWC 2024~\cite{conf}.}
\thanks{
The authors are with the Department of Electrical Engineering (ISY), Link\"oping University, 58183 Link\"oping, Sweden (email: jianan.bai@liu.se, anuch87@liu.se, anders.g.hansson@liu.se, erik.g.larsson@liu.se). This work was supported in part by the Excellence Center at Link\"oping-Lund in Information Technology~(ELLIIT), and by the Knut and Alice Wallenberg~(KAW) foundation.
}}
	
\maketitle

\begin{abstract}

	We consider a cellular massive MIMO system where swarms of wireless repeaters are deployed to improve coverage.
	These repeaters are full-duplex relays with small form factors that receive and instantaneously retransmit signals. 
	They can be deployed in a plug-and-play manner at low cost, while being  transparent to the network---conceptually they are  \emph{active channel scatterers} with amplification capabilities.
	Two fundamental questions need to be addressed in repeater deployments: 
	(i)~How can we prevent  destructive effects of positive feedback caused by inter-repeater interaction (i.e., each repeater receives and amplifies signals from others)?
	(ii)~How much performance improvement can be achieved given  that  repeaters also inject noise and may introduce more interference?
	To answer these questions, we first derive a generalized Nyquist stability criterion for the repeater swarm system, and provide an easy-to-check stability condition.
	Then, we study the uplink performance and develop an efficient iterative algorithm that  jointly optimizes the repeater gains, user transmit powers, and receive combining weights to maximize the weighted sum rate while ensuring system stability.
	Numerical results corroborate our theoretical findings and show that the repeaters can significantly improve the system performance, both in sub-6 GHz and millimeter-wave bands.  
	The results also warrant careful deployment to fully realize the benefits of repeaters, for example, by ensuring a high probability of line-of-sight links between repeaters and the base station.
\end{abstract}


\section{Introduction}

Massive \gls{mimo}, a key enabler for 5G cellular systems, provides
high spectral efficiency and allows multiple users to be
simultaneously served with low-complexity linear processing by
deploying large antenna arrays at the
\glspl{bs}~\cite{larsson2014massive}.  However, cellular massive
\gls{mimo} still suffers from poor cell-edge coverage due to severe
signal attenuation and multi-cell interference.  Additionally, the
propagation environment in cellular networks is complex and there
inevitably exist coverage holes due to shadowing and blockage.
Furthermore, when supporting multi-antenna users, the channel rank
deficiency, resulted from limited scattering, restricts the spatial
multiplexing gains of massive \gls{mimo}.

A potential solution is distributed \gls{mimo}, also known as
cell-free massive \gls{mimo}, where the antennas are grouped into many
\glspl{ap}, densely distributed across the coverage
area~\cite{ngo2017cell}.  In distributed \gls{mimo} systems, the
likelihood of a user being in the vicinity of some \glspl{ap} is
significantly increased (a benefit known as macro diversity),
effectively resolving the aforementioned issues.  However, a
widespread deployment of distributed \gls{mimo} is not yet practically
viable, primarily due to the demanding requirements for
high-capacity fronthaul, reciprocity calibration, and synchronization.

The limitation of cellular massive \gls{mimo} and the deployment
challenges of distributed \gls{mimo} call for a transitional paradigm
that retains the key benefits of distributed
\gls{mimo} while minimizing deployment overhead.  One promising
solution is the repeater swarm-assisted cellular massive MIMO system
conceptualized in \cite{willhammar2024achieving}, where large numbers
(swarms) of repeaters are deployed within the cells to assist the
signal propagation between the massive \gls{mimo} \glspl{bs} and the
users.  These repeaters are essentially one type of full-duplex relays, amplifying
and instantaneously retransmitting signals with minimal delay (less
than a microsecond), and they can have very small form factors.

The use of repeaters as such is not a new concept---its commercial
progress was initiated in 2G in order to improve coverage, especially
in tunnels, and has been considered for various scenarios over the
years~\cite{patwary2005capacity,sharma2015repeater,garcia2007enhanced,tsai2010capacity,ma2015channel,10848174,10409170}.
In 5G \gls{nr}, significant efforts have been made to standardize
\glspl{ncr} in 3GPP \cite{TS38-106}, enabling more functionalities through control signaling.  
The potential of \glspl{ncr} has been demonstrated in many recent
studies~\cite{carvalho2024network, kapuruhamy2025understanding,
  da2024impact, wen2024shaping, ayoubi2023network, leone2022towards,
  aastrom2024ris, da2024cellular, aghazadeh2024advanced,
  ayoubi2024optimal, guo2022comparison}.  In this paper, we adopt the
paradigm envisioned in \cite{willhammar2024achieving}, which differs
slightly from the aforementioned works on multi-antenna \glspl{ncr}:
we consider simple single-antenna repeaters without beamforming
capabilities, deployed in large numbers, densely within the cells.
Conceptually, these repeaters function as \emph{active channel
scatterers} that amplify the signal.  They can be deployed in a
plug-and-play manner at very low cost, in a manner transparent to both
the users and the \glspl{bs}.
Very few studies are available on the analysis and optimization of repeater swarms; an exception is  \cite{topal2025fair} that considers repeater gain optimization for max-min fairness and energy efficiency; however, inter-repeater interaction (to be explained shortly) and user power control are not taken into account therein.

Meanwhile, several practical aspects of repeater deployment require
careful consideration: in TDD operation they must be reciprocity
calibrated \cite{larsson2024reciprocity}; they should have a
\gls{lti} response; be band-selective; offer enough self-interference
mitigation; and have short delay.  We refer to
\cite{willhammar2024achieving} for a detailed discussion of these
properties.  

\subsection{Two Fundamental Aspects}

Most existing works on repeater-assisted cellular systems focus on
numerical studies, lacking a comprehensive theoretical analysis.
Additionally, inter-repeater interaction, a key factor that can cause
system instability, is neglected in all literature we are aware of
(except \cite{conf}).  With this motivation, in this paper we
investigate two fundamental aspects of repeater swarm-assisted
cellular systems:

\textbf{Interaction Stability:} As repeaters operate in
full-duplex, they inevitably pick up, amplify, and retransmit signals
from each other. This creates a positive feedback loop within the
repeater swarm, and this feedback loop can become unstable.  (For
analogy, imagine connecting a microphone to a speaker via an
amplifier, and placing the microphone near the speaker: the microphone
captures the sound from the speaker, amplifies and feeds it back to
the speaker, causing a loud screech.)  Theoretically, instability
means unbounded growth of output power or energy; in practice, it
leads to amplifier saturation and system malfunction.  To prevent such
issues, it is critical to identify conditions for system stability.
To this end, we analyze repeater swarms from a linear system-theoretic
perspective, and establish criteria for stability.

\textbf{Communication Performance:} Repeaters are not only signal
amplifiers, but also active noise sources; the injected noise will be
amplified along with the desired signals and received by the
destination node.  Meanwhile, as single-antenna devices without
beamforming capabilities, repeaters cannot spatially separate
signals. Especially in the uplink, signals from different users will
be mixed at the repeaters, potentially causing severe
interference. The performance of repeater swarm-assisted cellular
systems is determined by a complex interplay between the signal
amplification, noise injection, and interference mitigation.  We
develop a framework for quantifying the performance gains brought by
repeater swarms in realistic scenarios.

\subsection{Contributions and Organization of the Paper}

\begin{enumerate}[leftmargin=*]
	\item 
	In Section \ref{sec: system model}, we introduce the system
        and signal model of a repeater swarm-assisted massive
        \gls{mimo} system.  Particularly, the inter-repeater
        interaction is characterized and incorporated in this model.

	\item
	We analyze the stability of repeater swarm systems in
        Section \ref{sec: stability}.  Specifically, we derive a
        generalized Nyquist stability criterion for the system, and
        provide two sufficient conditions that are easy to verify in
        practice.  The effectiveness of the derived conditions is then
        demonstrated through several examples.

	\item 
	The uplink performance of   repeater swarm-assisted massive
        \gls{mimo} system is studied in Section \ref{sec:
          performance}, in terms of the sum capacity and  
        achievable rate under linear combining.  We
        develop an efficient iterative algorithm to jointly optimize the repeater gains,
        user transmit powers, and receive combining weights, to
        maximize the weighted sum rate while ensuring system stability.

	\item
	We present numerical results in Section \ref{sec:
          simulations}, illustrating the performance gains brought by
        repeaters in both sub-6 GHz and millimeter-wave bands.  We
        also provide practical insights into repeater deployment.

\end{enumerate}

This paper is a comprehensive extension of our conference paper
\cite{conf}, which contained a special case of the interaction
stability analysis. Herein we develop a general framework for
performance analysis, extend the stability analysis to arbitrary
amplification gains, and provide a rigorous proof of the stability
criterion stated in \cite{conf}.

\subsubsection*{Notation}

Time functions are written in Helvetica font, $\sfx(t)$, and Laplace
transforms in Italic font, $x(s)$.  Vectors and matrices are written
in boldface lowercase and uppercase, $\bx$ and $\bX$,
respectively. The determinant of $\bX$ is denoted $\det(\bX)$.
$(\cdot)^\transp$, $(\cdot)^\herm$, and $(\cdot)^{-1}$ denote
transpose, Hermitian (conjugate transpose), and inverse, respectively.
$\bI_N$ and $\bzero_N$ denote the identity matrix and all-zero vector
of size $N$ (omitted when no confusion can occur). $\E[\cdot]$ denotes
statistical expectation.  $\Re\{\cdot\}$ denotes the real part.  The
multivariate circularly symmetric complex Gaussian distribution with
covariance $\bR$ is denoted by $\cn{\bzero}{\bR}$. $\bD_{\bx}$ denotes
a diagonal matrix with $\bx$ on its diagonal.
$\R$ and $\C$ denote the spaces of real and complex numbers,
respectively, and $\C_+ \defeq \{s\in\C: \Re\{s\}\geq 0\}$ the complex
right half-plane.  $\abs{\cdot}$, $\norm{\cdot}$, and
$\matnorm{\cdot}$ denote absolute value, vector- or operator norm, and
matrix norm, respectively.

\section{Motivating Example}
\label{sec: example}

\begin{figure}
	\centering
	\includegraphics[width=0.35\textwidth]{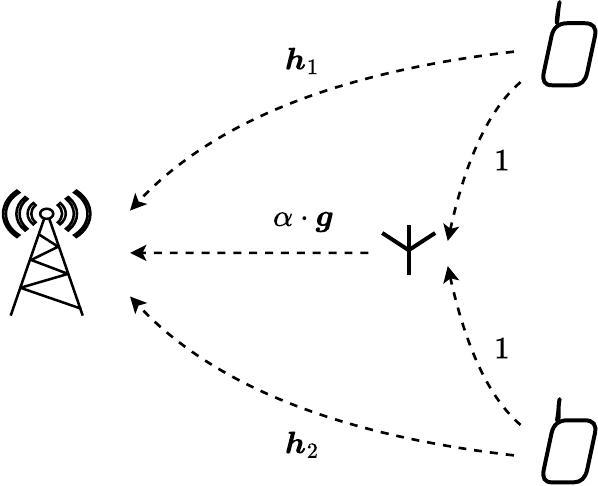}
	\caption{A motivating example.}
	\label{fig: example}
\end{figure}

Before investigating more complex systems, we first consider a simple setup with one \gls{bs}, one repeater, and two users, as illustrated in Fig. \ref{fig: example}. 
The repeater is placed equidistant from both users.
Intuitively, this is an unfavorable scenario, as the users cause strong interference to each other at the repeater.

\subsection{Best-Case Scenario with Orthogonal Channels}

The direct-link channels (i.e., user-to-\gls{bs}) are denoted by $\bh_1$ and $\bh_2$, and the \gls{r2b} channel is denoted by $\bg$. 
All the channels are normalized to have unit gain, i.e., $\norm{\bh_1} = \norm{\bh_2} = \norm{\bg} = 1$, and be mutually orthogonal, i.e., $\bh_1^\herm \bh_2 = \bh_1^\herm \bg = \bh_2^\herm \bg = 0$.
The orthogonality assumption represents the \emph{best-case} scenario in which the information over different channels can be perfectly separated at the \gls{bs} via linear processing.
The repeater has an amplification gain $\alpha \geq 0$.
The received signal at the \gls{bs} is given by
\begin{equation}
	\label{eq: example-received-signal}
	\by = \bh_1 x_1 + \bh_2 x_2 + \alpha \bg (x_1 + x_2) + \bw,
\end{equation}
where $x_1$ and $x_2$ represent the signals from the users, and $\bw \sim \cn{\bzero}{\varsigma^2 \bI}$ represent the \gls{awgn} at the \gls{bs} with variance  $\varsigma^2$.

Since the channels are orthogonal, we can obtain the sufficient statistics for decoding $x_1$ and $x_2$ by projecting the received signal $\by$ onto the directions of $\bh_1$, $\bh_2$, $\bg$:
\begin{equation}
	\left\{
	\begin{array}{l}
		\tilde{y}_1 = \bh_1^\herm \by = x_1 + \tilde{w}_1 \\
		\tilde{y}_2 = \bh_2^\herm \by = x_2 + \tilde{w}_2 \\
		\tilde{y}_3 = \bg^\herm \by = \alpha (x_1 + x_2) + \tilde{w}_3
	\end{array}
	\right.
\end{equation}
which can be written as the linear model
\begin{equation}
	\tilde{\by} 
	= 
	\begin{bmatrix}
		\tilde{y}_1 \\ \tilde{y}_2 \\ \tilde{y}_3
	\end{bmatrix}
	=
	\underbrace{
	\begin{bmatrix}
		1 & 0 \\
		0 & 1 \\
		\alpha & \alpha
	\end{bmatrix}
	}_{\displaystyle \defeq \bH}
	\begin{bmatrix}
		x_1 \\ x_2 
	\end{bmatrix}
	+
	\tilde{\bw},
\end{equation}
where $\bw \sim \cn{\bzero}{\varsigma^2 \bI_3}$.
We consider the best linear unbiased estimator (BLUE) [equivalently, the \gls{zf} combiner] of $x_1$ and $x_2$, given by
\begin{equation}
	\begin{bmatrix}
	\hat{x}_1\\ \hat{x}_2
	\end{bmatrix}
	 = (\bH^\transp \bH)^{-1} \bH^\transp \tilde{\by},
\end{equation}
which has error covariance $\varsigma^2 (\bH^\transp \bH)^{-1}$; therefore, the variance of each estimate is
\begin{equation}
	\label{eq: example-snr-gain}
	\frac{1 + \alpha^2}{1 + 2\alpha^2} \varsigma^2 \xrightarrow{\alpha \rightarrow \infty} \frac{1}{2} \varsigma^2.
\end{equation} 

We note that even in this idealized scenario, where channels are orthogonal and the repeater adds no noise, a poorly placed repeater can only offer at most 3 dB \gls{snr} gain, at the cost of infinitely large amplification.

\subsection{Non-Ideal Scenarios}

\begin{figure}
	\centering
	\includegraphics[width=0.45\textwidth]{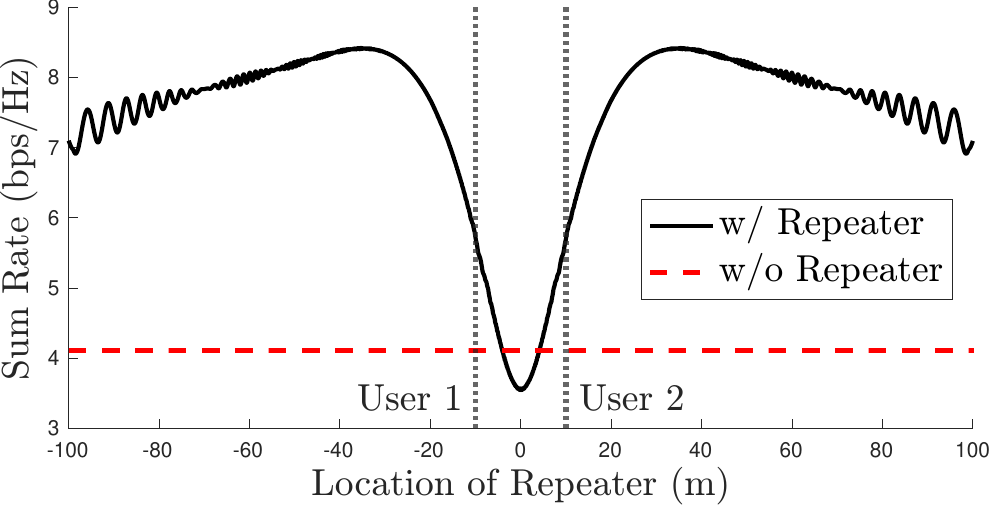}
	\caption{Averaged sum rate for different repeater locations.}
	\label{fig: example-simulation}
\end{figure}

We illustrate the limitation of a poorly placed repeater without proper user-repeater coordination by numerical results.   Fig.~\ref{fig: example-simulation} shows the performance
for a scenario where two users are placed 40 meters apart and 500 meters from the \gls{bs}.
The repeater is moved along a line parallel to, and 40 meters offset from, the line connecting the two users.
Both the users and the repeater have a maximum transmit power of 23 dBm, and the repeater amplification gain is constrained to be less than 90 dB.
The pathlosses and fading follow the 3GPP models that will be described in Section \ref{sec: simulations} at 6 GHz.
The \gls{bs} uses a \gls{mmse} combiner to decode the users' signals, and the sum rate is obtained by a moving average over the repeater locations for every 4 meters.
Notice that since the channels are no longer orthogonal, the {linear} \gls{mmse} combiner cannot completely eliminate the interference between different channels.
As we can observe, a repeater placed in the middle of the two users can negatively affect the sum rate due to increased interference.
This example illustrates the importance of joint control of the repeater gains and user transmit powers, as well as the need for a systematic analysis of repeater swarm-assisted cellular systems to fully understand the benefits and limitations of deploying swarms of repeaters.

\section{System Model}
\label{sec: system model}

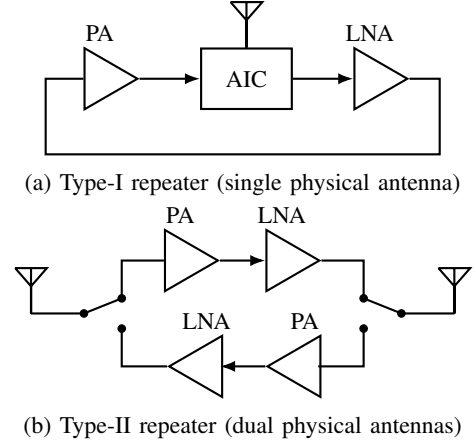
\begin{figure}
	\centering
	\begin{subfigure}[b]{0.45\textwidth}
		\centering
		\begin{tikzpicture}[
			>={Latex[length=2mm]},
			every node/.style={font=\small}
		]

		\node[amp, label={[yshift=2pt]above:PA}] (PA) at (0,0) {};

		\node[block, right=0.8cm of PA] (AIC) {AIC};

		\pic[scale=0.6] at (AIC.north) {antenna};
							
		\node[amp, right=0.8cm of AIC, label={[yshift=2pt]above:LNA}] (LNA) {};

		\draw[->, thick] (PA) -- (AIC.west);
		\draw[->, thick] (AIC.east) -- (LNA);

		\coordinate (taptarget) at ([xshift=-0.5cm,yshift=-1cm]PA.base west);
		\draw[thick] (LNA.base east) -- ++(0.4,0) -- ++(0,-1) -- (taptarget) -- ++(0,1) -- (PA.base west);

		\end{tikzpicture}
		\caption{Type-I repeater (single physical antenna)}
	\end{subfigure}
	\hfill
	\begin{subfigure}[b]{0.45\textwidth}
		\centering
		\begin{tikzpicture}[
		>={Latex[length=2mm]},
		every node/.style={font=\small}
		]
	
		\coordinate (ant1pos) at (0,0);
		\pic[scale=0.6] (ant1) at (ant1pos) {antenna} ;
		
		\coordinate (PA1pos) at (2.5,0.7);
		\node[amp, label={[yshift=2pt]above:PA}, at={(PA1pos)}] (PA1) {};
		\node[amp, label={[yshift=2pt]above:LNA}, right=0.6cm of PA1pos] (LNA1) {};
		
		\pic[scale=0.6, at={(5.6,0)}] (ant2) {antenna};
		
		\coordinate (LNA2pos) at ([xshift=1.8cm, yshift=-0.7cm] ant1pos);
		\node[amp, xscale=-1, label={[yshift=2pt]above:LNA}, at={(LNA2pos)}] (LNA2) {};
		\node[amp, xscale=-1, label={[yshift=2pt]above:PA}, at={(3.1,-0.7)}] (PA2) {};
		
		\node[circ, at={(0.7,0)}] {};
		\node[circ, at={(1.2,0.2)}] {};
		\node[circ, at={(1.2,-0.2)}] {};
		
		\node[circ, at={(4.9,0)}] {};
		\node[circ, at={(4.4,0.2)}] {};
		\node[circ, at={(4.4,-0.2)}] {};
		
		\draw[thick] (0,0) -- (0.7,0) -- (0.7,0) -- (1.2,0.2) -- (1.2,0.7) -- (PA1.base west) ;
		\draw[thick, ->] (PA1.base east) -- (LNA1.base west);
		\draw[thick] (LNA1.base east) -- (4.4,0.7) -- (4.4,0.2) -- (4.9,0) -- (5.6,0);
		\draw[thick] (1.2,-0.2) -- (1.2,-0.7) -- (LNA2.base east);
		\draw[thick, ->] (3.1,-0.7) -- (2.5,-0.7);
		\draw[thick] (3.8,-0.7) -- (4.4,-0.7) -- (4.4,-0.2);
		
	\end{tikzpicture}
		\caption{Type-II repeater (dual physical antennas)}
		\label{subfig: placement downlink}
	\end{subfigure}
	\caption{Block diagrams of repeater (redrawn from \cite{willhammar2024achieving}).}
	\label{fig: repeater diagram}
\end{figure}

\begin{figure*}
	\centering
	\includegraphics[width=0.8\textwidth]{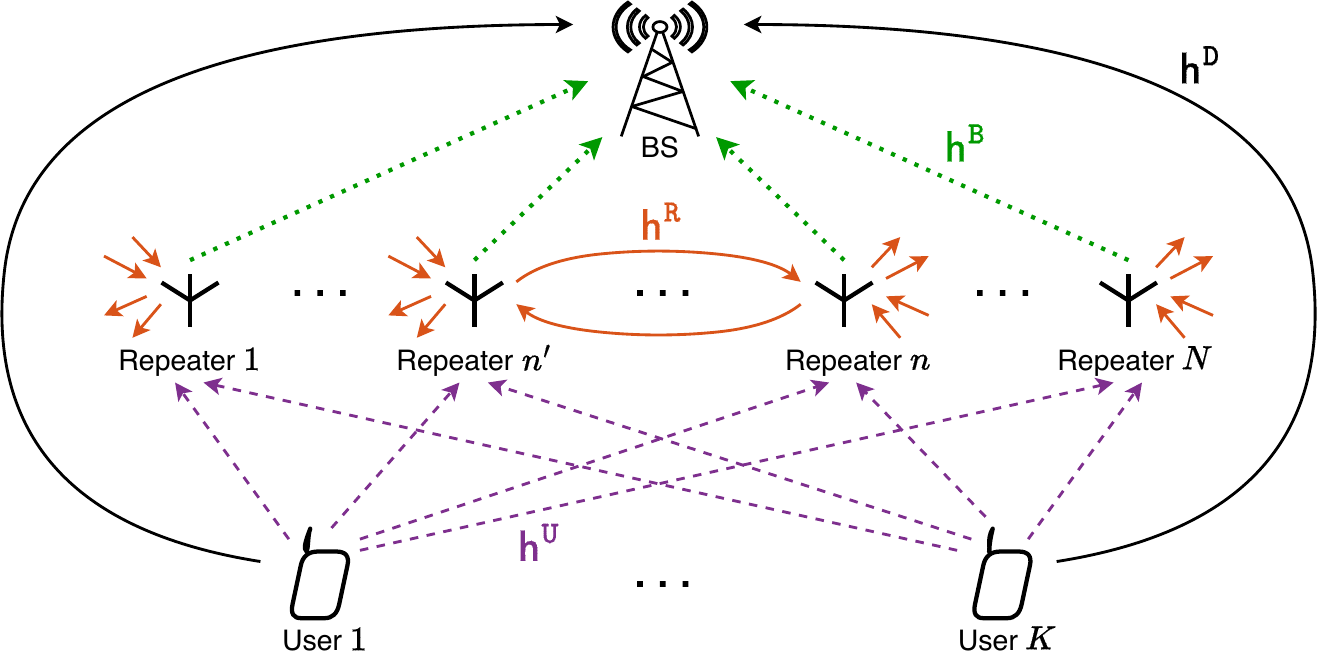}
	\caption{Illustration of a repeater swarm-assisted cellular system. }
	\label{fig: system model}
\end{figure*}

We consider a single-cell system where an $M$-antenna \gls{bs} serves $K$ single-antenna users. 
To improve coverage, $N$ repeaters are deployed across the cell. 

The repeaters are full-duplex relays with single antenna port, but they can have one or dual physical antennas, as shown in Fig. \ref{fig: repeater diagram}.
Type-I repeaters use the same physical antenna for reception and transmission. 
The received signal is first forwarded to a low-noise amplifier (LNA) and then to a power amplifier (PA). 
The self-interference is mitigated by an antenna-interface circuit (AIC), which prevents the amplified signal from the PA to re-enter the LNA.
Type-II repeaters employ two physical antennas, one pointing toward the \gls{bs} and another to the users.
A switch is used to select the transmission direction.
Compared with type-I repeaters, type-II repeaters can achieve better self-interference cancellation and one antenna can be properly tuned toward the \gls{bs} for better link quality.
However, type-II repeaters must be carefully calibrated for reciprocity-based communication \cite{larsson2024reciprocity}.

Each repeater, say repeater $n$, has the impulse response
\begin{equation}
	\label{eq: repeater impulse response}
	\begin{aligned}
		\sfa_n(t) = \alpha_n \delta(t-\nu_n),
	\end{aligned}
\end{equation}
where $\alpha_n \geq 0$ is the \emph{amplification gain}, and $\nu_n > 0$ represents the induced delay.
All wireless channels are assumed to be \gls{lti}\footnote{In practice, all wireless channels are time-varying. The assumption of a \gls{lti} system, is, strictly speaking, an approximation. However, since the transmission slot is typically much shorter than the channel coherence time, during which the channel response is nearly time-invariant, approximating the system as \gls{lti} is reasonable. Analyzing a time-varying system is interesting but much more challenging. We will have to leave it as a future direction. \label{footnote: LTI}} and reciprocal.
We use the following notations for the channels of different links (see Fig. \ref{fig: system model}):\footnote{These are all passband signals and \emph{not} complex baseband.}
\begin{center}	{\renewcommand{\arraystretch}{2.3}%
	\begin{tabular}{|ll|}
		\hline
		$\sfh^\ttD_{mk}(t)$ & \makecell[l]{\underline{\textbf{D}}irect Link:\\ user $k$ $\leftrightarrow$ $m$th BS antenna}\\[5pt]
		\hline
		$\sfh^{\ttU}_{nk}(t)$ & \makecell[l]{\underline{\textbf{U}}ser-Repeater Link: \\ user $k$ $\leftrightarrow$ repeater $n$}\\[5pt]
		\hline
		$\sfh^\ttR_{nn'}(t)$ & \makecell[l]{Inter-\underline{\textbf{R}}epeater Link:\\ repeater $n$ $\leftrightarrow$ repeater $n'$}\\[5pt]
		\hline
		$\sfh^{\ttB}_{mn}(t)$ & \makecell[l]{\underline{\textbf{B}}S-Repeater Link:\\ repeater $n$ $\leftrightarrow$ $m$th BS antenna}\\[5pt]
		\hline
	\end{tabular}
	}
\end{center}

Since we will have to deal with potentially unstable systems, for which the Fourier transform may not exist, we will work in the Laplace domain at this stage.
For a function $\sfx(t)$ defined for $t \geq 0$, the Laplace transform is expressed as a function of the complex variable $s$:
\begin{equation}
	x(s) \defeq \int_0^\infty \sfx(t) e^{-st} dt. 
\end{equation}
For example, the repeater impulse response $\sfa(t)$ in \eqref{eq: repeater impulse response} has the following Laplace transform in the region of convergence $\C_+$:
\begin{equation}
\label{eq: repeater transfer function}
	a_n(s) = \alpha_n e^{-s \nu_n}.
\end{equation}

The Laplace transforms are written in Italic font, $x(s)$, to distinguish the corresponding time functions in Helvetica font, $\sfx(t)$.
When the Fourier transform exists, we can represent it by choosing $s = j\omega$, where $\omega$ is the angular frequency.
Once the stability conditions are established in Section \ref{sec: stability}, we will transition to analysis in the frequency domain.

We next derive the input-output relationship of the repeater swarms, which is used for stability and performance analysis.

\subsection{The Repeater Swarm}

The repeaters operate in full-duplex, instantaneously amplifying and retransmitting the received signals.
Let $u_n(s)$ denote the input to repeater $n$, consisting of signals from the intended sources (i.e., user signals in uplink or BS signal in downlink) and \gls{awgn}. 
The output from repeater $n$ is given by
\begin{equation}
	\label{eq: single repeater output}
	\begin{aligned}
		r_{n}(s) = a_{n}(s) \bigg( u_n(s)
		+ \sum_{n'=1}^N h^\ttR_{nn'}(s) r_{n'}(s) \bigg),
	\end{aligned}
\end{equation}
where the summand corresponds to the repeater interaction for $n'\neq n$ and self-interference for $n'=n$.

By concatenating \eqref{eq: single repeater output} for all repeaters, the output from the repeater swarm, $\br(s) \defeq [r_1(s),\cdots,r_N(s)]^\transp$, satisfies
\begin{equation}
\label{eq: repeater output equation}
	\br(s) = \bD_{\ba}(s) \big(\bu(s) + \bH^\ttR(s) \br(s) \big),
\end{equation}
where 
$\ba(s) \defeq [a_{1}(s),\cdots,a_N(s)]^\transp$, $\bu(s) \defeq [u_1(s),\cdots,u_N(s)]^\transp$, and
\begin{equation}
    \bH^\ttR(s) \defeq 
	\begin{bmatrix}
        h_{11}^\ttR(s) & \cdots & h^\ttR_{1N}(s)\\
        \vdots & \ddots & \vdots\\
        h^\ttR_{N1}(s) & \cdots & h^\ttR_{NN}(s)
    \end{bmatrix}.
\end{equation}
We note that $\bH^\ttR(s)$ is symmetric due to channel reciprocity; however, it is not Hermitian in general.

\begin{figure}
	\centering
	\includegraphics[width=0.35\textwidth]{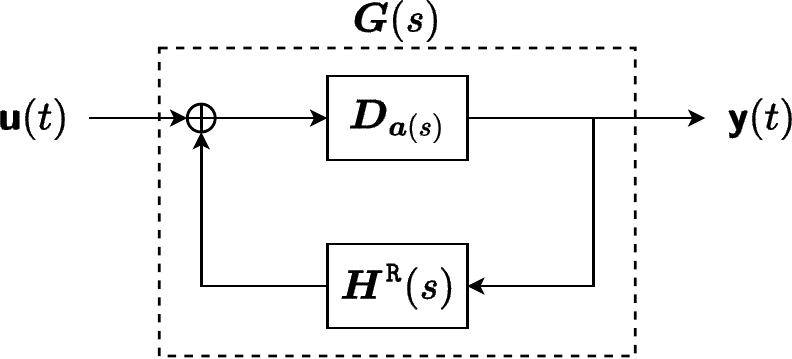}
	\caption{Block diagram of the positive-feedback system.}
	\label{fig: block diagram}
\end{figure}

By rearranging \eqref{eq: repeater output equation}, the input-output relation of the repeater swarm is expressed as
\begin{equation}
\label{eq: repeater swarm output equation}
    \br(s) = \bG(s) \bu(s),
\end{equation}
where
\begin{equation}
	\label{eq: effective response of repeaters}
	\bG(s) \defeq \big(\bI_N - \bD_{\ba(s)}\bH^\ttR(s) \big)^{-1}\bD_{\ba(s)}
\end{equation}
is the effective transfer function matrix of the repeater swarm.
$\bG(s)$ corresponds to a multi-dimensional system with a positive feedback loop due to repeater interaction, as illustrated by the block diagram in Fig. \ref{fig: block diagram}.
Note that $\bG(s) = \bG(s)^\transp$ but $\bG(s) \neq \bG(s)^\herm$.

We next present the uplink and downlink signal models of the repeater-assisted massive \gls{mimo} system.

\subsection{Uplink Model}

Let $x_{k}^{\ul}(s)$ denote the signal transmitted by user $k$. 
Then, the input to the repeater swarm is 
$
	\bu^{\ul}(s) = \bH^\ttU(s) \bx^{\ul}(s) + \bw^{\ttR,\ul}(s),
$
where $\bx^{\text{ul}}(s) \defeq [x_{1}^\text{ul}(s),\cdots,x_{K}^\ul(s)]^\transp$, $\bH^\ttU(s) \in \C^{N\times K}$ contains the channel coefficients $\{h^\ttU_{nk}(s)\}$, and $\bw^{\ttR,\ul}(s) \in \C^N$ represents the \gls{awgn} at the repeaters.
The received signal at the \gls{bs} consists of both the direct-link signals from the users and the  signals from the repeaters:
\begin{align}
	\by^\ul(s)
	=&~ \bH^\ttD(s) \bx^{\ul}(s) + \bH^\ttB(s) \bG(s) \bu^\ul(s) + \bw^{\ttB}(s) \notag \\
	=&~ (\bH^\ttD(s)+ \bH^\ttB(s) \bG(s) \bH^\ttU(s)) \bx^{\ul}(s) \notag \\ 
	&+ \bH^\ttB(s) \bG(s) \bw^{\ttR,\ul}(s) + \bw^\ttB(s), 
	\label{eq: uplink model}
\end{align}
where 
$\bH^\ttD(s) \in \C^{M\times K}$ contains $\{h^\ttD_{mk}(s)\}$, $\bH^\ttB(s) \in \C^{M\times N}$ contains $\{h^\ttB_{mn}(s)\}$, and $\bw^{\ttB}(s) \in \C^M$ denotes the \gls{awgn} at the \gls{bs}.

\subsection{Downlink Model}
Let $\bx^{\text{dl}}(s) \in \C^M$ denote the precoded signal transmitted by the \gls{bs}.
Due to channel reciprocity, the uplink and downlink channel matrices are equal, up to a transpose.
Thus, the input to the repeater swarm is $\bu^\dl(s) = (\bH^\ttB)(s)^\transp \bx^\dl(s) + \bw^{\ttR,\dl}(s)$, where $\bw^{\ttR,\dl}(s) \in \C^N$ is the repeater noise.
The received signal at user $k$ is given by
\begin{align}
	y_k^\dl(s)
	=&~ \bh_k^\ttD(s)^\transp \bx^\dl(s) + \bh_k^\ttU(s)^\transp \bG(s) \bu^\dl(s) + w_k(s) \notag \\
	=&~ ( \bh_k^\ttD(s) + \bH^\ttB(s) \bG(s) \bh_k^\ttU(s) )^\transp \bx^\dl(s) \notag\\
	&+ (\bG(s) \bh_k^\ttU(s))^\transp \bw^{\ttR,\dl}(s) + w_k(s), \label{eq: downlink model}
\end{align}
where $\bh_k^\ttD(s)$ and $\bh_k^\ttU(s)$ denote the $k$th columns of $\bH^\ttD(s)$ and $\bH^\ttU(s)$, respectively, and $w_k$ represents the \gls{awgn} at user $k$.

\section{Stability Analysis}
\label{sec: stability}

Repeater interactions create a positive feedback loop, requiring careful configuration to prevent instability.
Stability of the system is fully determined by the \emph{joint} response of 
the repeater swarm and is \emph{agnostic} to whether the system operates in the uplink or downlink.
In this section, we investigate the conditions for system stability.

\subsection{Bounded-Energy Stability}

Before it becomes relevant to analyze stability of the system, it is necessary to first verify that the system is well-defined---specifically, that the transfer function matrix $\bG(s)$ represents the Laplace transform of a \emph{unique} and \emph{causal} impulse response $\mat{\sfG}(t)$.
We will formally address this point in Theorem~\ref{th: stability}.
For now, assuming the system is well-defined and $\mat{\sfG}(t)$ exists, we introduce the bounded-energy stability criterion.
The relevant definitions are given as follows.

\begin{definition}
	A time signal $\vect{\sfu}(t)$ of arbitrary dimension is said to have finite energy if $\int_{-\infty}^{+\infty} \vect{\sfu}(t)^\transp \vect{\sfu}(t) \mathrm{d}t < \infty$.
\end{definition}

\begin{definition}
	An \gls{lti} system with a causal impulse response $\mat{\sfG}(t)$ of arbitrary dimension is bounded energy stable if, for any input signal $\vect{\sfu}(t)$  with finite energy, the output signal $\int_0^t \mat{\sfG}(\tau) \vect{\sfu}(t-\tau) d\tau$ has finite energy.
\end{definition}

\begin{remark}
	Bounded-energy stability ensures that the system remains stable for bounded-energy inputs.
	The system is bounded-energy stable if its transfer function has a finite $\cH_\infty$ norm \cite[Th. 4.3]{zhou1998essentials}.
\end{remark}

\begin{remark}
	\label{remark: stability}
	Although stability has been extensively studied in control theory, most existing results are derived under the assumption that the system has a rational transfer function; see, for example, the Nyquist stability criterion in \cite[Th.~4.7]{skogestad2005multivariable}.
	The presence of time delays---which are inherent in communication systems---results in infinite-dimensional state spaces that do not have a rational transfer function.
	While it is generally true that the Nyquist stability criterion applies to non-rational systems as well, a rigorous proof 
	of this fact is involved; see, for instance, \cite{chait1989nyquist}.
	(In \cite{chait1989nyquist}, bounded-input bounded-output stability is considered only for one-dimensional systems, rather than the bounded-energy stability for multi-dimensional systems we consider here.)
\end{remark}

To establish our main results on stability, we introduce the following two assumptions.

\begin{assumption}
	\label{as: analyticity}
	The transfer function matrix $\bH^\ttR(s)$ is analytic in the right half-plane $\C_+$.\footnote{Strictly speaking, a function is considered analytic only in open sets. 
	Therefore, when we say a function is analytic in $\C_+$, we mean there exists a $\gamma>0$ such that the function is analytic in $\{s \in \C: \Re(s) > -\gamma\}$.}
\end{assumption}

\begin{assumption}
\label{as: channel amplitude gain}
	There exist constants $C,\varepsilon,\delta>0$ such that the inter-repeater channel amplitude gain satisfies	
	$|h_{nn'}^\ttR(s)| \leq \frac{C}{|s|^{1+\varepsilon}} $ for all $|s| \geq \delta$ in $\C_+$ for all $n,n'\in \{1,2,\cdots,N\}$.
\end{assumption}

\begin{remark}
	Physically, Assumption \ref{as: channel amplitude gain} implies that the channel amplitude decays with frequency at a rate faster than $1/|s|$ for large $|s|$.
	This reflects practical wireless propagation behavior: in free space, the channel amplitude already scales as $1 / |\omega|$ due to the decreasing effective aperture of antennas, and in real environments, due to lossy media or material penetration, the attenuation is even faster.
	Additionally, antennas and RF front-ends inherently act as bandpass filters, limiting the response at high frequencies.
\end{remark}

We now give our main results in the following theorem.

\begin{theorem}
	\label{th: stability}
	Under Assumptions \ref{as: analyticity} and \ref{as: channel amplitude gain}, if the image of $\det(\bI_N - \bD_{\ba(j\omega)} \bH^\ttR(j\omega))$ does not encircle the origin, then for $\bG(s)$ defined in \eqref{eq: effective response of repeaters}, the impulse response
	\begin{equation}
		\mat{\sfG}(t) \defeq \frac{1}{j2\pi} \int_{\sigma-j\infty}^{\sigma+j\infty} \bG(s) e^{s t}~ \mathrm{d} s
	\end{equation}
	exists for all $\sigma \geq 0$ and is independent of the choice of $\sigma$.
	Additionally, $\mat{\sfG}(t)$ is causal, i.e., $\mat{\sfG}(t) = \bzero$ for $t<0$, and bounded-energy stable.

\end{theorem}
\begin{IEEEproof}
		See Appendix.
	\end{IEEEproof}
\begin{remark}
	When the conditions in Theorem~\ref{th: stability} hold, the region of convergence of the Laplace transform includes the $j\omega$ axis; hence, the Fourier transform $\bG(j\omega)$ is well-defined, allowing us to analyze the system in the frequency domain.
\end{remark}

\begin{remark}
	Theorem \ref{th: stability}  generalizes  the Nyquist stability criterion in \cite[Th.~4.7]{skogestad2005multivariable} to the particular  multi-dimensional system that we study in this paper.
\end{remark}

\subsection{Simplified Condition for Interaction Stability}

Theorem \ref{th: stability} provides a stability condition expressed in terms of  the image of the complex-valued function $\det(\bI_N - \bD_{\ba(j\omega)} \bH^\ttR(j\omega))$. This function has a   complicated appearance in general (see example in Fig. \ref{fig: Nyquist plot for circle}), making the result difficult to apply in practice. 
To guide practical system design, we develop a \emph{sufficient}, more restrictive,  condition for stability, which
is much simpler and  depends explicitly on the channel gains between repeaters.
The key idea is to apply Gershgorin disc theorem to find the region, i.e., union of discs, where the eigenvalues of $\bI_N - \bD_{\ba(j\omega)} \bH^\ttR(j\omega)$ are located and make sure they are bounded away from zero. 

\begin{propi}
	\label{propi: simplified stability}
	The non-encirclement condition in Theorem \ref{th: stability} is satisfied if $\sup_{\omega \in \R} D(\balpha;\omega) < 1$, where
    \begin{equation}
        D(\balpha;\omega) \defeq \min\left\{D_1(\balpha;\omega), D_2(\balpha;\omega)\right\},
    \end{equation}
    with
	\begin{subequations}
		\begin{alignat}{2}
			D_1(\balpha;\omega) &\defeq \max_n ~\alpha_n \sum_{n'=1}^N \abs{h_{nn'}^\ttR(j\omega)},\\
			D_2(\balpha;\omega) &\defeq \max_n \sum_{n'=1}^N \alpha_{n'} \abs{h_{nn'}^\ttR(j\omega)}.
		\end{alignat}
	\end{subequations}
\end{propi}

\begin{IEEEproof}
	By applying the Gershgorin disc theorem \cite[Th. 6.1.1]{horn2012matrix} to both the rows and columns of $\bD_{\ba(j\omega)}\bH^\ttR(j\omega)$, it follows that for an arbitrary $\omega$, all eigenvalues of $\bD_{\ba(j\omega)}\bH^\ttR(j\omega)$ lie within the intersection of the Gershgorin sets:
	\begin{subequations}
		\begin{alignat*}{1}
		&~ \bigcup_{n} \left\{ z \in \C : \abs{z - a_n h_{nn}^\ttR(j\omega)} \leq \alpha_n \sum_{n'\neq n} \abs{h_{nn'}^\ttR(j\omega)} \right\},\\
		&~ \bigcup_{n} \left\{ z \in \C : \abs{z - a_n h_{nn}^\ttR(j\omega)} \leq \sum_{n'\neq n} \alpha_{n'} \abs{h_{nn'}^\ttR(j\omega)} \right\},
		\end{alignat*}
	\end{subequations}
	which are enclosed by the circles centered at the origin with radii $D_1(\balpha;\omega)$ and $D_2(\balpha;\omega)$, respectively.
	Therefore, their intersection lies within the circle centered at the origin with radius $D(\ba;\omega)$.
	(The use of Gershgorin disc theorem in relation to multivariable Nyquist techniques was also exploited in, for example, \cite[Ch. 2.10]{maciejowski1989multivariable}.)

	To proceed, notice that $D(\balpha;\omega)$ is  a monotonically non-decreasing function of $\balpha$, i.e., $D(\balpha';\omega) \geq D(\balpha;\omega)$ for any $\balpha'$ such that $\alpha_n' \geq \alpha_n, \forall n$.
	When all repeaters are turned off, i.e., $\balpha_n=\bzero$, the image of $\det(\bI_N - \bD_{\ba(j\omega)} \bH^\ttR(j\omega))$ collapses to a single point at $1+j0$, and the system is trivially stable.
	As the amplification gains continuously increase from zero, the image of $\det(\bI_N - \bD_{\ba(j\omega)} \bH^\ttR(j\omega))$ will continuously change, and the system will remain stable as long as the image has not yet intersected the origin. 
	The non-intersection of that image with the origin can be guaranteed if $D(\balpha;\omega) < 1$ for all $\omega$.	
\end{IEEEproof}

\begin{remark}
	While checking the conditions for Proposition \ref{propi: simplified stability} requires evaluating the inter-repeater channel gains for all frequencies, this can be simplified in practice.
	Since wireless systems operate within a limited frequency range and antennas inherently function as bandpass filters, it is generally sufficient to verify the condition within the operational frequency range of interest (e.g., 20~MHz bandwidth at 6~GHz and 100~MHz at 30~GHz, 60~GHz, and 70~GHz \cite[Table 7.8-2]{TR38-901}).
    Alternatively, one could get a looser bound by taking the maximum over $\omega$ for each $h_{nn'}^\ttR(j\omega)$ term inside the expression of $D(\balpha,\omega)$---for each inter-repeater link, we only need to know the maximum channel gain over the operating frequency band.
	Nevertheless, it is worth noting that spurious emissions, amplifier harmonics, and wideband interference or jamming may introduce out-of-band behavior that could, in some cases, affect the system. A wider-spectrum analysis could offer further insights in scenarios where such effects become non-negligible.
\end{remark}

\subsection{Special Cases}

To gain further insights, we consider the special case where all repeaters have the same amplification gain $\alpha$.

\begin{corollary}
	\label{coro: simplified stability}
	For the special case when $a_n(s) = \alpha e^{-s\nu_n}, \forall n$, the non-encirclement condition in Theorem \ref{th: stability} is satisfied if
	\begin{equation}
		\alpha < \alpha_G \defeq \inf_\omega \min_n \frac{1}{\sum_{n'=1}^N |h_{nn'}^\ttR(j\omega)|}. 
	\end{equation}
\end{corollary}

\begin{remark}
	Corollary \ref{coro: simplified stability} was presented in \cite{conf} when ignoring the time-delay.\footnote{In \cite{conf} it was erroneously mentioned in the concluding remarks that as a possible extension of the analysis,  $\alpha$ could be complex-valued. But it cannot be since $\alpha$ applies to the passband signal. What was intended by that remark in \cite{conf} is that a time-delay of the repeater could be included in the model, which we have done herein.}
	The Gershgorin theorem analysis is the same but \cite{conf} postulated without proof that stability holds if $\bI - \alpha' \bH^\ttR(j\omega)$ is non-singular for all $\omega$ and $\alpha' \leq \alpha$;  we prove this rigorously here.
	\end{remark}

\begin{remark}
	For stability of the repeater swarm, it is the sum of channel \emph{amplitude} gains that matters, rather than the sum of channel \emph{power} gains.
	In the worst case, the positive feedback combines constructively, in-phase, behaving as if the repeaters formed a coherent antenna array.
\end{remark}

Next, we consider two examples with \gls{los} channels and ignore the channel gain variation with frequency (narrow enough bandwidth of the signal), such that
\begin{equation}
	h_{nn'}^\ttR(s)  = \sqrt{\beta_{nn'}^\ttR} e^{-s \tau_{nn'}},
\end{equation} 
where $\beta_{nn'}^\ttR$ is the channel power gain and $\tau_{nn'}$ is the propagation delay that are both determined by the distance $d_{nn'}$.
The condition in Corollary \ref{coro: simplified stability} simplifies to
\begin{equation}
	\alpha < \alpha_G = \min_n \frac{1}{\sum_{n'\neq n} \sqrt{\beta_{nn'}^\ttR}}.
\end{equation}
For simplicity, we also consider that all repeaters have the same delay $\nu$, such that $a_n(s) = a(s) = \alpha e^{-s\nu}, \forall n$.

\subsubsection{Two Repeaters}

Consider the case of $N=2$ repeaters. 
We can write for simplicity $h_{12}(s) = h_{21}(s) = \sqrt{\beta} e^{-s\tau}$, and the inter-repeater channel transfer function matrix is
\begin{equation}
	\bH^\ttR(s) = 
	\begin{bmatrix}
		0 & \sqrt{\beta} e^{-s\tau} \\
		\sqrt{\beta} e^{-s\tau} & 0
	\end{bmatrix};
\end{equation}
therefore,
\begin{equation}
	\det(\bI_2 - a(s) \bH^\ttR(s)) = 1 - \alpha^2 \beta e^{-2s(\tau + \nu)}.
\end{equation}
The image of $\det(\bI_2 - a(j\omega) \bH^\ttR(j\omega))$ traces out a circle centered at $1+j0$ with radius $\alpha^2\beta$ anticlockwise and periodically for every change of $\omega$ by $\pi / (\tau + \nu)$.
According to Theorem~\ref{th: stability}, the system is stable if $\alpha^2\beta < 1$.

For this particular case, the following recursive equations can be written in the time domain:
\begin{subequations}
	\begin{alignat}{2}
		\sfr_1(t) &= \alpha \sfu_1(t-\nu) + \alpha \sqrt{\beta}~ \sfr_2(t-\tau-\nu),\\
		\sfr_2(t) &= \alpha \sfu_2(t-\nu) + \alpha \sqrt{\beta}~ \sfr_1(t-\tau-\nu),
	\end{alignat}
\end{subequations}
which yield the output of the two repeaters as
\begin{equation*}
	\begin{aligned}
		\sfr_1(t) &= \alpha \sfu_1(t-\nu) * \sfp(t) + \alpha^2 \sqrt{\beta}~ \sfu_2(t-\nu) * \sfp(t-\tau-\nu), \\
		\sfr_2(t) &= \alpha \sfu_2(t-\nu) * \sfp(t) + \alpha^2 \sqrt{\beta}~ \sfu_1(t-\nu) * \sfp(t-\tau-\nu).
	\end{aligned}
\end{equation*}
Here, $\sfp(t)$ represents the impulse train
\begin{equation}
	\sfp(t) = \sum_{i=0}^{\infty} (\alpha^2\beta)^i\delta(t - 2i(\tau + \nu)),
\end{equation}
which represents the ``ping-pong'' effect of the loopback signal between the two repeaters.
The system is stable if $\alpha^2\beta < 1$, ensuring that the impulse train decays exponentially to zero.
It is noteworthy that both Theorem \ref{th: stability} and Proposition \ref{propi: simplified stability} provide necessary and sufficient conditions for this special case.

\subsubsection{Repeaters on Circle}

\begin{figure}
	\centering
	\begin{tikzpicture}[scale=0.7]
  	\draw[dashed, thick] (0,0) circle (2cm);
  
 	\pgfmathsetmacro{\anglestep}{360/7}
  	\pgfmathsetmacro{\sinSexty}{sin(50)}
  	\pgfmathsetmacro{\sinThirty}{sin(40)}
  
 	 \foreach \i in {0,1,2,3,4,5,6}
    {
      \pgfmathsetmacro{\angle}{\i*\anglestep+90}
      \draw[fill] (\angle:2cm) circle (2.5pt) node[anchor=center] (dot\i) {};
      \ifnum\i=0
        \coordinate (firstDot) at (\angle:2cm);
      \fi
      \ifnum\i=1
        \coordinate (secondDot) at (\angle:2cm);
      \fi
      \ifnum\i=5
        \coordinate (fourthDot) at (\angle:2cm);
      \fi
    }
    \draw[thick] (firstDot) -- (secondDot) node[midway, below] {$d_1$};
    \draw[thick] (firstDot) -- (fourthDot) node[midway, right] {$d_2$};
    
  	\draw[thick] (-0.15*\sinSexty,0.15*\sinThirty) -- (0,0); 
  	\draw[thick] (0.15*\sinSexty,0.15*\sinThirty) -- (0,0); 
  	\draw[thick] (0,-0.15) -- (0,0); 
  
	\end{tikzpicture}
	\caption{Repeaters equally spaced on a circle.}
	\label{fig: circular network}
\end{figure}
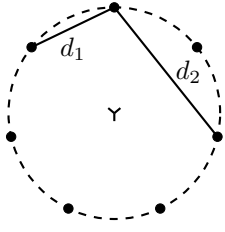

Consider an odd number of repeaters, $N=2N_0+1$, for some integer $N_0$, uniformly spread over a circle with radius $R$. 
(The analysis carries over to an even number of repeaters with slight changes, omitted here for brevity.)
Next, with a slight abuse of notation, we define 
\begin{equation}
	d_i \defeq 2R \sin(i\pi/N),
\end{equation}
and denote the channel power gain and the delay at distance $d_i$ as $\beta_i$ and $\tau_i$, respectively. 
By indexing the repeaters in the clockwise order, the channel between repeaters $n$ and $n'$ is
\begin{equation}
\label{eq: k-hop los channel}
	h_i^\ttR(s) \defeq \sqrt{\beta_i} e^{-s \tau_i},
\end{equation}
with
\begin{equation}
	i = \min\{|n-n'|,N-|n-n'|\}.
\end{equation}
See Fig. \ref{fig: circular network} for a graphical illustration.

In this particular case, $\bH^\ttR(s)$ is symmetric circulant with each column being a circulant permutation of the vector
\begin{equation}
	\bq(s) \defeq [0, h_1^\ttR(s),\cdots, h_{N_0}^\ttR(s), h_{N_0}^\ttR(s), \cdots, h_1^\ttR(s)]^\transp.
\end{equation}
We can now compute all eigenvalues of $\bH^\ttR(j\omega)$ as the \gls{dft} of $\bq(j\omega)$, and obtain
\begin{alignat}{2}
	\label{eq: det for circle}
	&\det(\bI_N - a(j\omega) \bH^\ttR(j\omega)) \nonumber \\
	&=\! \prod_{n=1}^{N}\!\! \left(1 \!-\! 2 \alpha \sum_{i=1}^{N_0} \left(\!\!\sqrt{\beta_i}\cos\! \frac{2\pi i (n\!-\!1)}{N}\right) e^{-j\omega(\tau_i + \nu)}\right).
\end{alignat}
Although \eqref{eq: det for circle} provides a closed-form expression, determining the shape of the image of $\det(\bI - a(j\omega) \bH^\ttR(j\omega))$ and whether it encircles the origin for an arbitrary $\alpha$ remains challenging. 

\begin{figure}
	\centering
	\begin{subfigure}[b]{6.9cm}
		\includegraphics[width=\textwidth]{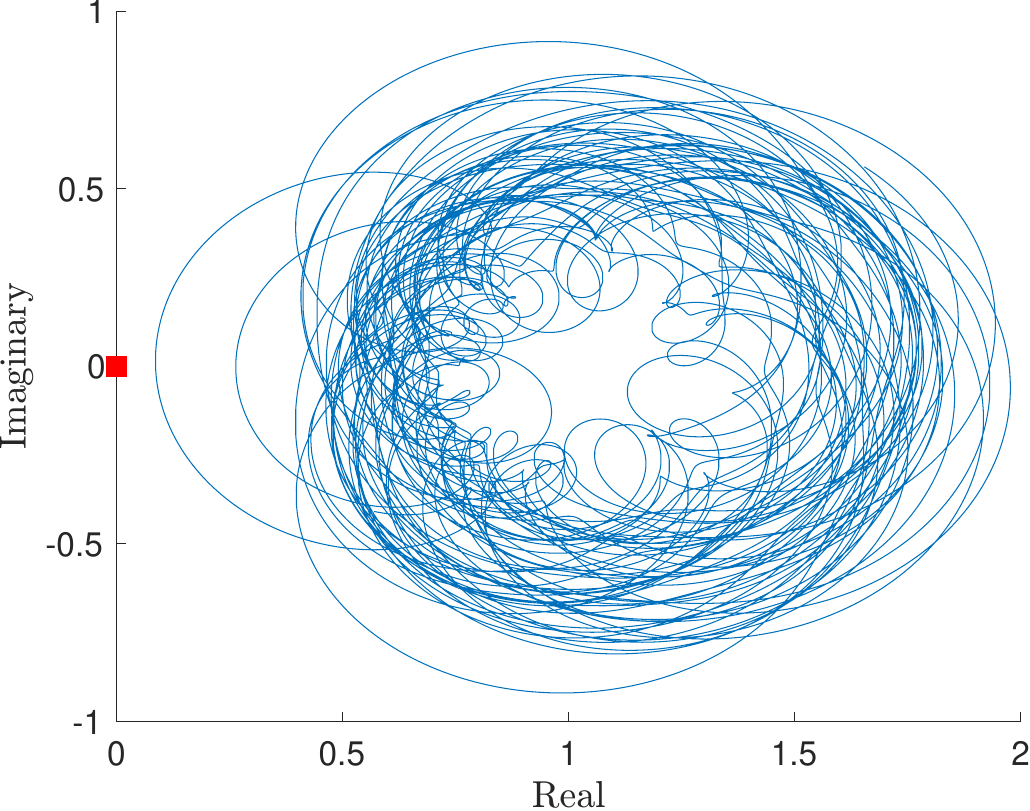}
		\caption{The image of $\det(\bI_N - \alpha \bH^\ttR(j\omega))$ with the amplification gain $\alpha=\alpha_G \approx 75.8$ dB}
		\label{fig: Nyquist plot for circle}
	\end{subfigure}
	\par\medskip
	\begin{subfigure}[b]{7.5cm}
		\includegraphics[width=\textwidth]{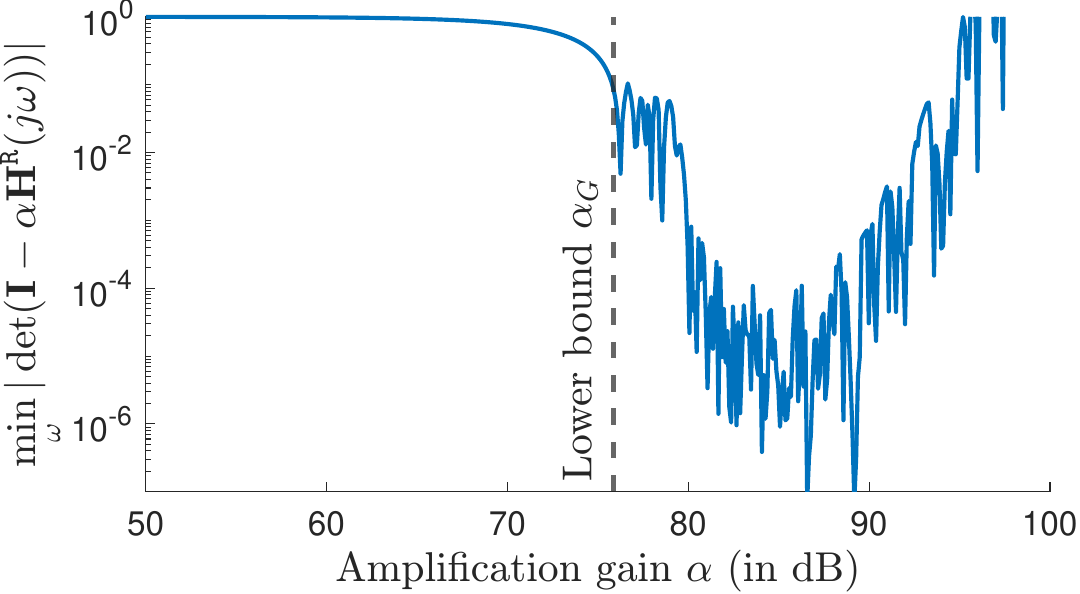}
		\caption{Plot of $\displaystyle \min_{\omega}\abs{\det(\bI_N - \alpha \bH^\ttR(j\omega))}$ for different $\alpha$}
		\label{fig: frequency sweeping for circle}
	\end{subfigure}
	\caption{Numerical results when 15 repeaters are uniformly spaced on a circle with a radius of 1000 meters.}
	\label{fig: policy}
\end{figure}

We visualize the image of $\det(\bI - \alpha \bH^\ttR(j\omega))$ in Fig.~\ref{fig: Nyquist plot for circle} under the free-space propagation model $\beta_i = c^2 / (2 \omega d_i)^2$ and $\tau_i = d_i / c$.
The repeater amplification gains are selected as the critical value $\alpha_G$, which guarantees interaction stability according to Corollary \ref{coro: simplified stability}.
The samples are taken by frequency sweeping over a 20~MHz band centered at 2~GHz, with a step size of 100~Hz.
Notice that even in this special case, appearance of the image is quite complex.
Additionally, in Fig. \ref{fig: frequency sweeping for circle}, we plot the minimum value of $\abs{\det(\bI_N - \alpha \bH^\ttR(j\omega))}$ over the sampled frequency points for different values of $\alpha$.
It can be observed that the critical value $\alpha_G$ accurately captures the transition point at which the system starts to become unstable.

\subsection{How much of the instability effect will be seen within a coherence time interval?}
\label{sec: instability effect}

For a finite-dimensional causal system with a proper rational transfer function, the stability is determined by the poles of the transfer function. 
If all poles are in the left half-plane (therefore have negative real parts), the system is stable; after triggered by an impulse, the system response will converge to a zero   after a transient time.
The order of magnitude of the transient time is determined by the pole closest to the imaginary axis, say $p$, which corresponds to an output that exponentially decays with the rate $e^{-|\Re(p)| t}$.

For  an unstable system of finite dimension, one can make a rational fraction expansion of the transfer function. The term with the pole furthest into the right half  plane will eventually dominate over   the other terms, and determines the
instability behavior as $t$ grows.  

However, since we have a system with time-delays, it  has  infinite dimension and does not have a rational transfer function. Hence, the argument above does not directly apply. We have to leave a more accurate, quantitative investigation of
the instability behavior for future work.

\section{Uplink Performance}
\label{sec: performance}

In this section, we analyze  the communication performance of the repeater-assisted system. 
All the analyses are performed in the frequency domain, within the coherence bandwidth centered at an arbitrary carrier (angular) frequency $\omega$.
Hence, the received signal follows the uplink model in \eqref{eq: uplink model} and the downlink model in \eqref{eq: downlink model}, with $s=j\omega$. 
For brevity, we focus on the uplink; once the uplink performance is determined, one can analyze or optimize the downlink by exploiting uplink-downlink duality \cite{shi2007downlink}.
We also note that, as observed in \cite{da2024impact}, uplink communications benefit more from repeaters due to the limited transmit power of user devices.

\begin{remark}
	We consider a single-carrier, narrowband system for ease of exposition. However, all  results can be extended to multi-carrier (wideband) systems.
	In principle, the optimization for a multi-carrier system has the same problem structure as \eqref{P1}
	presented below---it introduces an outer sum over the sub-carriers in the objective function, and the same solution techniques can be applied. 
	The optimization variables (combining vectors, repeater gains, and user powers) can either be kept the same for all sub-carriers or optimized independently for each sub-carrier.
\end{remark}

We first rewrite the received signal in \eqref{eq: uplink model} as
\begin{equation}
	\by(j\omega) = \bH(j\omega) \bx(j\omega) + \bw(j\omega),
\end{equation}
where the composite uplink channel, $\bH(j\omega)$, and aggregate noise at the \gls{bs}, $\bw(j\omega)$, are respectively expressed as
\begin{subequations}
	\begin{alignat}{2}
		\bH(j\omega) \defeq&~ \bH^\ttD(j\omega) + \bH^\ttB(j\omega) \bG(j\omega) \bH^\ttU(j\omega), \label{eq: composite channel} \\
		\bw(j\omega) \defeq&~ \bw^{\ttB}(j\omega) + \bH^\ttB(j\omega) \bG(j\omega) \bw^{\ttR}(j\omega).
	\end{alignat}
\end{subequations}
With this model, we first analyze the uplink capacity and then develop a linear combining scheme to maximize the weighted sum rate through joint repeater gain configuration and user power control under the interaction stability constraint.

For clarity, the explicit dependence on $j\omega$ and the superscript $(\cdot)^\ul$ will be omitted henceforth. It is worth noting that performance metrics such as capacity, \gls{sinr}, and achievable rates may, in general, vary with frequency.

\subsection{Capacity Analysis}
\label{sec: capacity analysis}

Let $\bw^\ttB \sim \cn{\bzero}{\varsigma_\ttB^2 \bI_M}$ and $\bw^\ttR \sim \cn{\bzero}{\varsigma_\ttR^2 \bI_N}$, where $\varsigma_\ttB^2$ and $\varsigma_\ttR^2$ are the noise powers at the \gls{bs} and repeaters, respectively. 
The aggregate noise $\bw$ is therefore colored Gaussian, i.e., $\bw \sim \cn{\bzero}{\bSigma}$, with   covariance matrix
\begin{equation}
\label{eq: noise covariance}
	\bSigma = \varsigma_\ttB^2 \bI_M + \varsigma_\ttR^2 \bH^\ttB \bG \bG^\herm (\bH^\ttB)^\herm.
\end{equation}
We can pre-whiten the received signal as
\begin{equation}
\label{eq: pre-whitening}
	\bSigma^{-\frac{1}{2}} \by = \bSigma^{-\frac{1}{2}} \bH \bx + \bSigma^{-\frac{1}{2}} \bw,
\end{equation}
where $\bSigma^{-\frac{1}{2}} \bw \sim \cn{\bzero}{\bI_M}$.

The pre-whitened signal model in \eqref{eq: pre-whitening} represents a \gls{mimo} multiple-access channel with \gls{awgn}.
Considering the power constraint $\E[|x_k|^2] \leq P_{\max}$ for each user, the sum capacity of the system is given by \cite{varanasi1997optimum}
\begin{align}
	\label{eq: sum capacity}
	C_{\mathtt{sum}}
	=& \log \det\left(\bI_M + P_{\max}\bSigma^{-\frac{1}{2}} \bH \bH^\herm \bSigma^{-\frac{1}{2}} \right) \notag \\
	=& \log \det\left(\bI_M + P_{\max} \bSigma^{-1} \bH \bH^\herm \right).
\end{align}

Let $\{R_k\}_{k=1}^{K}$ denote the achievable rates of all users. 
The uplink capacity region of the system can be expressed as \cite{varanasi1997optimum}
\begin{equation}
\begin{aligned}
	\Bigg\{
	&(R_1,\cdots,R_K): \forall \cS \subset \{1,2,\cdots,K\}, \\
	&\sum_{k\in\cS} R_k \leq \log\det\left(\bI_M + P_{\max} \bSigma^{-1} \sum_{k\in\cS} \bh_k \bh_k^\herm  \right) 
	\Bigg\},
\end{aligned}
\end{equation}
where $\bh_k$ denotes the $k$th column of $\bH$.

\subsection{Linear Combining \& Weighted Sum-Rate Maximization}
\label{sec: linear combining}

Achieving the sum capacity in \eqref{eq: sum capacity} requires the use of \gls{sic}, which is nonlinear and computationally difficult in practice.
We instead consider linear combining---the \gls{bs} uses a vector $\bc_k \in \C^M$ to combine the received signal so as to decode data from user $k$.
The signal after combining is 
\begin{equation}
\begin{aligned}
	\hat{q}_k 
	\defeq&~ \bc_k^\herm \by 
	= \bc_k^\herm \bH \bx + \bc_k^\herm \bw \\
	=&~ \sqrt{\rho_k}\bc_k^\herm \bh_k q_k + \underbrace{\sum_{k' \neq k} \sqrt{\rho_{k'}} \bc_k^\herm \bh_{k'} q_{k'} + \bc_k^\herm \bw}_{\text{interference-plus-noise}},
\end{aligned}
\end{equation}
where we express the transmit signal as $x_n \defeq \sqrt{\rho_k} q_k$; here, $q_k$ is the data symbol of user $k$ with $\E[|q_k|^2] = 1$, and $\rho_k \in [0,P_{\max}]$ denotes the transmit power.

Considering uncorrelated data symbols, the instantaneous \gls{sinr} of user $k$ is
\begin{equation}
	\sinr_k = \frac{\rho_k \bc_k^\herm \bh_k \bh_k^\herm \bc_k}{\bc_k^\herm\Bigg(\displaystyle \sum_{k' \neq k} \rho_{k'} \bh_{k'}\bh_{k'}^\herm + \bSigma \Bigg)\bc_k}, \label{eq: SINR_k}
\end{equation}
which is a generalized Rayleigh quotient with respect to $\bc_k$, and can be maximized by choosing \gls{mmse} combiner\footnote{
Notice that the scaling factor is canceled out in the ratio and is therefore irrelevant for \gls{sinr} or rate maximization. We chose the particular scaling because it also minimizes the \gls{mse}.}
\begin{equation}
	\label{eq: mmse combiner}
	\bc_k^\mmse = \sqrt{\rho_k} \left(\bH \bD_{\brho} \bH^\herm + \bSigma\right)^{-1} \bh_k.
\end{equation}
The resulting \gls{sinr} with the \gls{mmse} combiner becomes
\begin{alignat}{2}
	\label{eq: mmse sinr}
	\sinr_k^\mmse =&~ \rho_k \bh_k^\herm \left(\sum_{k'\neq k} \rho_{k'} \bh_{k'} \bh_{k'}^\herm + \bSigma\right)^{-1} \bh_k \notag \\
	=&~ \frac{1}{1 - \rho_k \bh_k^\herm \left(\bH \bD_{\brho} \bH^\herm + \bSigma\right)^{-1} \bh_k} - 1,
\end{alignat} 
where the equality follows from \cite[Lemma B.4]{bjornson2017massive}.

Now, the weighted sum-rate maximization problem under the repeater stability constraint can be formulated as
\begin{alignat}{2}
	&\maximize_{\{\bc_k\},\brho,\balpha} &&~ \sum_{k=1}^K \gamma_k \log\left(1 + \sinr_k \right) \tag{P1} \label{P1}\\
	&~~~~\subjectto &&~ 0 \leq \rho_k \leq P_{\max}, ~\forall k \tag{C1} \label{C1} \\
	& &&~ 0 \leq \alpha_n \leq A_{\max}, ~\forall n \tag{C2} \label{C2} \\
	& &&
	\left\{
    \begin{aligned}  
        & \alpha_n \sum_{n'=1}^N \abs{h_{nn'}^\ttR} \leq \eta, ~\forall n, \mbox{ \underline{\textbf{or}}} \\  
        & \sum_{n'=1}^N \alpha_{n'} \abs{h_{nn'}^\ttR} \leq \eta, ~\forall n 
    \end{aligned}  
    \right.
	\tag{C3} \label{C3} 
\end{alignat}
where $\brho$ and $\balpha$ denote the collections of all $\{\rho_k\}$ and $\{\alpha_n\}$, respectively, $\gamma_k \geq 0$ is the weight that represents the priority of user $k$, $A_{\max}$ is the maximum repeater amplification gain  which should be set considering the isolation between input- and output antennas to avoid self-oscillation, and $\eta \in (0,1]$ is a pre-determined parameter that ensures that the repeaters operate away from the instability boundary (additionally, $\eta$ can provide a margin for robust stability under channel uncertainties).
Constraints \eqref{C3} is derived from the interaction stability requirements in Corollary \ref{coro: simplified stability}. 
Notice that either one of the conditions in \eqref{C3} is sufficient for stability according to Proposition \ref{propi: simplified stability}, and they are both linear constraints.

\textbf{Simplification $\bG = \bD_{\ba}$:} 
The repeater frequency responses, $\ba$, are absorbed---in a complicated way---in the matrix $\bG = (\bI - \bD_{\ba} \bH^\ttR)^{-1} \bD_\ba$, which appears in the composite channel matrix $\bH$ and the noise covariance $\bSigma$.  
For tractability, we ignore the inter-repeater interaction and consider $\bG = \bD_{\ba}$ when solving the optimization problem.
This simplification may lead to sub-optimality, as the model adopted for optimization is mismatched from the true model.
However, by imposing the stability constraint \eqref{C3}, the inter-repeater interaction should be kept small, making $\bG \approx \bD_{\ba}$ a reasonable approximation for balancing complexity against performance.
The resulting  scheme can be seen as a ``partial \gls{mmse} combiner'', where the inter-repeater interference is not accounted for in the combining vector design, to reduce complexity and avoid the need for full inter-repeater \gls{csi}.

\begin{remark}
\label{remark: time-delay}
	The repeaters add time-delays $\{\nu_n\}$, which, as shown in \eqref{eq: repeater transfer function}, introduce \emph{frequency-dependent} phase shifts $\{e^{-j\omega \nu_n}\}$ in the frequency responses.
	While all of our analyses account for arbitrary phase shifts, we restrict the optimization to the real-valued amplification gains $\balpha$.
	In principle, it is possible for the repeaters to actively adjust time-delays to control the phase shifts $\{e^{-j \omega \nu_n}\}$, so that the signals from different repeaters can combine more constructively at the BS for a specific frequency $\omega$.
	However, this would significantly complicate the optimization problem, increase the operational complexity of the repeaters, and make the system more sensitive to instantaneous channel variations.
	Such complexity is undesirable, as the repeaters are intended to be simple, low-cost devices that require infrequent reconfiguration to maintain maximum transparency to the network.
	We will leave the optimization of time-delays for possible future work. 
\end{remark}

The objective function of \eqref{P1} is non-convex, making it difficult to solve. 
To circumvent this problem, we use the fact that the weighted sum-rate maximization problem has the same optimal solution as a weighted \gls{mmse} problem  \cite[Th. 1]{shi2011iteratively}.
To see this, we first write the \gls{mse} of user $k$ as
\begin{alignat}{2}
	\xi_k 
	\defeq&~ \E\left[|\hat{q}_k - q_k|^2\right] \notag \\
	=&~ \bc_k^\herm  \left(\bH \bD_{\brho} \bH^\herm + \bSigma \right) \bc_k - 2 \sqrt{\rho_k} \Re\left\{\bc_k^\herm \bh_k\right\} + 1. \label{eq: mse}
\end{alignat}
The weighted \gls{mmse} problem is then formulated as
\begin{alignat}{2}
	&\minimize_{\bvarpi,\{\bc_k\},\brho,\balpha} &&~ \sum_{k=1}^K \gamma_k \left( \varpi_k  \xi_k - \log \varpi_k \right) \tag{P2} \label{P2}\\
	&~~~~\subjectto &&~ \eqref{C1}, \eqref{C2}, \eqref{C3}, \notag \\
	& &&~ \varpi_k \geq 0, ~\forall k, \tag{C4} \label{C4}
\end{alignat}
where a new variable $\bvarpi = [\varpi_1,\cdots,\varpi_K]^\transp$ is introduced to represent the \gls{mse} weights.
Optimizing $\bvarpi$ when fixing $\brho$, $\balpha$, and $\{\bc_k\}$ gives $\varpi_k^\opt = 1/\xi_k$. 
After substituting $\varpi_k^\opt$ back, the new optimization objective (after removing irrelevant constant terms) is to minimize $\sum_{k=1}^K \gamma_k \log \xi_k$, and the optimal $\{\bc_k\}$ when fixing $\brho$ and $\balpha$ are the \gls{mmse} combiners in \eqref{eq: mmse combiner}. 
Upon substituting $\{\bc_k^\mmse\}$, the optimization objective becomes minimizing $\sum_{k=1}^K \gamma_k \log \xi_k^\mmse$, where 
\begin{align}
	\label{eq: mmse mse}
	\xi_k^\mmse 
	&= 1 - \rho_k \bh_k^\herm \left( \bH \bD_{\brho} \bH^\herm + \bSigma \right)^{-1} \bh_k \notag \\
	&= 1 - \sqrt{\rho_k} \bh_k^\herm \bc_k^\mmse.
\end{align}
Comparing with \eqref{eq: mmse sinr}, one can observe $\log \xi_k^\mmse = -\log(1+\sinr_k^\mmse)$.
Clearly, \eqref{P1} and \eqref{P2} are equivalent and have the same optimal solution for $\{\bc_k\}$, $\brho$ and $\balpha$.

To solve the weighted \gls{mmse} problem in \eqref{P2}, we adopt a block coordinate descent algorithm similar to the one proposed in \cite{shi2011iteratively}, where we sequentially optimize one of the variables $\bvarpi,\{\bc_k\},\brho,\balpha$ while keeping the others fixed; as we will show, each sub-problem is convex.
As discussed above, the optimal $\bvarpi$ is given by $\varpi_k^\opt = 1/\xi_k^\mmse, \forall k$, where $\xi_k^\mmse$ is given in \eqref{eq: mmse mse}, and the optimal $\{\bc_k\}$ are the \gls{mmse} combiners in \eqref{eq: mmse combiner}.
We next derive the update rule for $\brho$ and $\balpha$.

\underline{\textit{Update Rule for} $\brho$}:
After substituting \eqref{eq: mse} into \eqref{P2} and omitting terms that do not depend on $\brho$, we can rewrite the objective function as\footnote{Notice that, when choosing $\bc_k$ as the \gls{mmse} combiner in \eqref{eq: mmse combiner}, $\bc_k^\herm \bh_k$ is positive-valued; thus, the $\Re\{\cdot\}$ operation can be omitted.} 
\begin{alignat}{2}
	& \sum_{k=1}^K \gamma_k \varpi_k \left( \sum_{k'=1}^K \rho_{k'} \abs{\bc_k^\herm \bh_{k'}}^2 - 2 \sqrt{\rho_k} \Re\left\{ \bc_k^\herm \bh_k \right\}  \right) \notag \\
	&= \sum_{k=1}^K \Bigg(\rho_k \sum_{k'=1}^K \gamma_{k'}\varpi_{k'}\abs{\bc_{k'}^\herm \bh_k}^2 - 2\sqrt{\rho_k} \gamma_k\varpi_k\Re\{\bc_k^\herm \bh_k\} \Bigg), \notag
\end{alignat}
which, along with the power constraint \eqref{C1}, decouples across users for different $\rho_k$.
The optimization of $\brho$ can thus be solved independently for each user $k$ as
\begin{alignat}{2}
	&\min_{\rho_k} &&~~ \rho_k \sum_{k'=1}^K \gamma_{k'}\varpi_{k'}\abs{\bc_{k'}^\herm \bh_k}^2\!\! - 2\sqrt{\rho_k}\gamma_k\varpi_k\Re\{\bc_k^\herm \bh_k\} \tag{S1} \label{S1} \\
	&~\subjectto &&~~~ 0 \leq \rho_k \leq P_{\max}  \notag
\end{alignat}
The objective function of \eqref{S1} is a one-dimensional quadratic function of $\sqrt{\rho_k}$, and the optimal solution is given by
\begin{equation}
	\label{eq: rho update}
	\rho_k^\opt = \min\left\{P_{\max}, \left(\frac{\gamma_k\varpi_k\Re\{\bc_k^\herm \bh_k\}}{\sum_{k'=1}^K \gamma_{k'}\varpi
	_{k'}\abs{\bc_{k'}^\herm \bh_k}^2}\right)^2 \right\}.
\end{equation}

\underline{\textit{Update Rule for} $\balpha$}:
Since we are only interested in optimizing the real-valued amplification gains $\balpha$, we can absorb the phase shifts into the \gls{bs}-repeater channels by defining $\widetilde{\bH}^\ttB$, whose columns are obtained by phase-shifting the corresponding columns in $\bH^\ttB$, i.e., $\tilde{\bh}_n^\ttB \defeq e^{j\omega \nu_n} \bh_n^\ttB $.
After the above simplification, we have $\bH = \bH^\ttD + \widetilde{\bH}^\ttB \bD_{\balpha} \bH^\ttU$ and $\bSigma = \bI + \varsigma^2 \widetilde{\bH}^\ttB \bD_{\balpha}^2 (\widetilde{\bH}^\ttB)^\herm$, substituting which into \eqref{eq: mse} we observe that the \gls{mse} $\xi_k$ is a convex quadratic function of $\balpha$.
Specifically, after some algebraic manipulation and omitting terms that do not depend on $\balpha$, we obtain
\begin{equation}
\label{eq: mse w.r.t. alpha}
	\xi_k \propto \balpha^\transp \bGamma_k \balpha + 2\bpsi_k^\transp \balpha, 
\end{equation}
where 
\begin{subequations}
\begin{alignat}{2}
	\bGamma_k &\defeq \Re\left\{ \bD_{\bphi_k}^\herm \left( \bH^\ttU \bD_{\brho} (\bH^\ttU)^\herm + \varsigma_\ttR^2 \bI \right) \bD_{\bphi_k} \right\} \\
	\bpsi_k &\defeq \Re\left\{ \bD_{\bphi_k}^\herm \left(\bH^\ttU \bD_{\brho} (\bH^\ttD)^\herm \bc_k - \sqrt{\rho_k} \bh_k^\ttU \right) \right\},
\end{alignat}
\end{subequations}
with $\bphi_k \defeq (\widetilde{\bH}^\ttB)^\herm \bc_k$.
We further define $\bGamma \defeq \sum_{k=1}^K \gamma_k \varpi_k \bGamma_k$ and $\bpsi \defeq \sum_{k=1}^K \gamma_k \varpi_k \bpsi_k$.
Notice that $\bGamma_k$ is \gls{psd}, so is $\bGamma$.
Now, the optimization of $\balpha$ can be cast as a convex \gls{qp} with linear constraints:
\begin{alignat}{2}
	&\min_{\balpha} &&~~~ \frac{1}{2}\balpha^\transp \bGamma \balpha + \bpsi^\transp \balpha \tag{S2} \label{S2} \\
	&~\subjectto &&~~~ \eqref{C2}, \eqref{C3}  \notag
\end{alignat}
which can be efficiently solved using standard toolboxes.

The overall optimization procedure is summarized in Algorithm \ref{alg: wmmse}.
Since the objective function of \eqref{P2} is continuously differentiable, and a unique minimizer can be found when updating each variable, convergence of the algorithm to a stationary point is guaranteed \cite[Prop. 2.7.1]{bertsekas1997nonlinear}.

\textbf{Repeater Power Constraint:}
In practice, the repeater gain is also constrained by its maximum power $P_{\max}^\ttR$ to keep  the PAs  working in their linear region and to satisfy energy consumption requirements.
From an implementation perspective, this can be enforced by incorporating the following  constraint while updating the repeater gains $\balpha$ by solving \eqref{S2}:
\begin{equation}
	\label{C7}
	\alpha_n^2\left( \sum_{k=1}^K \rho_k |h_{nk}^\ttU|^2 + \varsigma_\ttR^2\right) \leq  P_{\max}^\ttR, ~\forall n. \tag{C5}
\end{equation}
Notice that this constraint depends on the user powers $\brho$ which may change across iterations.
The objective value (i.e., weighted sum-rate) is still guaranteed to converge, as it is monotonically non-decreasing over the iterations, and upper-bounded. 
However, while the objective function is guaranteed to converge, the optimization variables are not (there could be  multiple feasible solutions that give the same objective function value).
Having said that, in our experimental results, we observe that the algorithm still converges well under this additional constraint.

\begin{algorithm}[t]
	\caption{Joint Uplink Optimization}
	\begin{algorithmic}[1]
		\label{alg: wmmse}
		\REQUIRE Channels $\bH^\ttD$, $\bH^\ttB$, $\bH^\ttU$; noise variances $\varsigma_\ttB^2$ and $\varsigma_\ttR^2$; user weights $\{\gamma_k\}$; maximum power $P_{\max}$; stopping threshold $\epsilon$; maximum number of iterations $I_{\max}$
		\INITIALIZE $\brho$ and $\balpha$ satisfying \eqref{C1}, \eqref{C2}, \eqref{C3}
		\WHILE{The change in $\sum_{k=1}^K \gamma_k \log \varpi_k$ exceeds $\epsilon$ \underline{\textbf{and}} the maximum iteration $I_{\max}$ is not reached}
		\STATE $\bc_k \gets \sqrt{\rho_k} \left(\bH \bD_{\brho} \bH^\herm + \bSigma\right)^{-1} \bh_k, \forall k$ \label{alg-line: mmse combiner}
		\STATE $\varpi_k \gets 1 / \big( 1 - \sqrt{\rho_k} \bh_k^\herm \bc_k \big), \forall k $
		\STATE $\displaystyle \rho_k \gets \min\left\{P_{\max}, \left(\frac{\gamma_k\varpi_k \bc_k^\herm \bh_k}{\sum_{k'=1}^K \gamma_{k'}\varpi
		_{k'}\abs{\bc_{k'}^\herm \bh_k}^2}\right)^2\right\}, \forall k$
		\vspace{0pt}
		\STATE Update $\balpha$ by solving the convex \gls{qp} \eqref{S2} \label{alg-line: repeater gain}
		\ENDWHILE
		\ENSURE $\{\bc_k\}$, $\brho$, and $\balpha$
	\end{algorithmic}
\end{algorithm}

\subsection{Signaling Aspects}
\label{sec: signaling aspects}

To perform the joint optimization of beamforming weights, user powers, and repeater gains, we need three types of \gls{csi} that involve repeaters: \gls{r2b} channels, \gls{u2r} channels, and \gls{r2r} channels.
For the \gls{r2r} channels, we \emph{do not} need full \gls{csi}---only their \emph{amplitudes} are needed to check the stability criterion in \eqref{C3}.
Furthermore, since the repeaters have fixed locations, the \gls{r2b} and \gls{r2r} channels will likely vary slowly over time compared to the direct-link and \gls{u2r} channels, and will therefore not need to be frequently estimated, implying that the overhead of \gls{r2b} and \gls{r2r} channel estimation does not impact every coherence block of data transmission.

\noindent $\triangleright$ \textbf{Estimation of slow-varying R2B and R2R channels:}

If the repeaters are capable of transmitting a pilot tone,
we can let the BS estimate the \gls{r2b} channels, and the other repeaters can estimate the \gls{r2r} channels.
Since the phase of the \gls{r2r} channel is not required, one could estimate the \emph{amplitudes} using envelope- or energy detectors.
This procedure requires an oscillator and an envelope-detector circuit at repeaters, but  baseband circuitry  is not needed.

\noindent $\triangleright$ \textbf{Estimation of fast-varying direct-link and U2R channels:}

As in conventional \gls{tdd} systems, we let the users transmit orthogonal pilot sequences to the BS in the uplink at the beginning of each coherence block.
If we ignore repeater interaction (which should be kept small to ensure stability once \gls{r2r} channel amplitudes are estimated), 
and ignore repeater noise for now, the received pilot signals at the BS, according to the uplink system model in \eqref{eq: uplink model}, can be written as
\begin{equation}
	\bY^\mathtt{p} = (\bH^\ttD + \bH^\ttB \bD_\ba \bH^\ttU) \bPhi + \bw,
\end{equation}
where $\bPhi$ is the pilot matrix with $\bPhi \bPhi^\herm = \bI_K$, and $\bw$ is the  noise at the BS.
After the estimation of R2B channels, $\bH^\ttB$ is known at the BS.
The users can perform pilot transmission twice, with the repeater gains set to $\ba$ and $-\ba$ (phase rotation of $\pi$), respectively.
Then, after de-spreading (i.e., calculating $\bY^\mathtt{p} \bPhi^\herm$), the BS can obtain two measurements
\begin{subequations}
\begin{alignat}{2}
	\bX^{(0)} &= \bH^\ttD + \bH^\ttB \bD_\ba \bH^\ttU + \bw^{(0)}, \\
	\bX^{(1)} &= \bH^\ttD - \bH^\ttB \bD_\ba \bH^\ttU + \bw^{(1)}.
\end{alignat}
\end{subequations}
By processing the measurements according to
\begin{subequations}
\begin{alignat}{2}
	\frac{\bX^{(0)} + \bX^{(1)}}{2} &= \bH^\ttD + \frac{\bw^{(0)} + \bw^{(1)}}{2}, \\
	\frac{\bX^{(0)} - \bX^{(1)}}{2} &= \bH^\ttB \bD_\ba \bH^\ttU + \frac{\bw^{(0)} - \bw^{(1)}}{2},
\end{alignat}
\end{subequations}
the noises remain uncorrelated, and the channels $\bH^\ttD$ and $\bH^\ttU$ can  be estimated  using, for example, least-squares or linear MMSE estimators.
These ideas are reminiscent of the techniques used
in \cite{larsson2024reciprocity}
for reciprocity calibration of repeaters.


We also note that separate estimates of individual channels are only needed when the joint optimization of beamforming weights, user transmit powers, and repeater gains is performed.
However, in practice, the repeater gains, as well as user transmit powers, mainly depend on large-scale fading, and can be updated at a much slower timescale than the beamforming weights at the BS (which need to be updated per coherence block).
To update the beamforming weights, we only need to know the \emph{effective} channel, $\bH \defeq \bH^\ttD + \bH^\ttB \bG \bH^\ttU$, which can be estimated directly using uplink pilots from users without involving the repeaters (which can be seen as active scatterers) in any special way.

We envision that practical operation of a repeater-assisted system would involve \gls{csi} acquisition on three timescales: 1) the effective channel $\bH$ is estimated every coherence block using uplink pilots from users to update the beamforming weights at the \gls{bs}; 2) the direct-link and \gls{u2r} channels are estimated when the large-scale fading changes significantly (e.g., every few seconds) to update the user powers and repeater gains; and 3) the \gls{r2b} channels and \gls{r2r} amplitudes are estimated even less frequently (e.g., every few tens of seconds).
One could also  consider deriving  optimization schemes that only require statistical \gls{csi} at the repeaters, perhaps relying on lower bounds on ergodic capacity.

Regarding the signaling overhead, the estimation of the effective channel $\bH$ requires $K$   uplink pilot symbols per coherence block, which is the same as in conventional massive \gls{mimo} systems without repeaters.
The estimation of direct-link and \gls{u2r} channels requires $2K$ uplink pilot symbols every few seconds.
The estimation of \gls{r2b} channels and \gls{r2r} channel amplitudes requires $N$ symbols of pilot transmission from the repeaters, and additionally $N$ symbols for feeding back the sum of \gls{r2r} channel amplitudes (i.e., $\{\sum_{n'=1}^N |h_{nn'}^\ttR|\}_{n=1}^N$) to the BS, every few tens of seconds.

\subsection{The Effects of Self-Interference}
\label{sec: self-interference}

\begin{figure}
	\centering
	\begin{subfigure}[b]{0.4\textwidth}
		\includegraphics[width=0.9\textwidth]{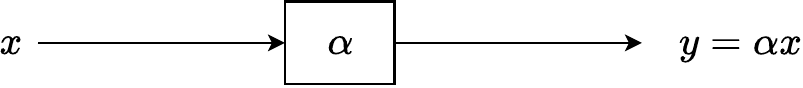}
		\caption{w/o self-interference}
		\label{subfig: wo-self-interference}
		\vspace{0.2cm}
	\end{subfigure}
	\hfill
	\begin{subfigure}[b]{0.4\textwidth}
		\centering
		\includegraphics[width=\textwidth]{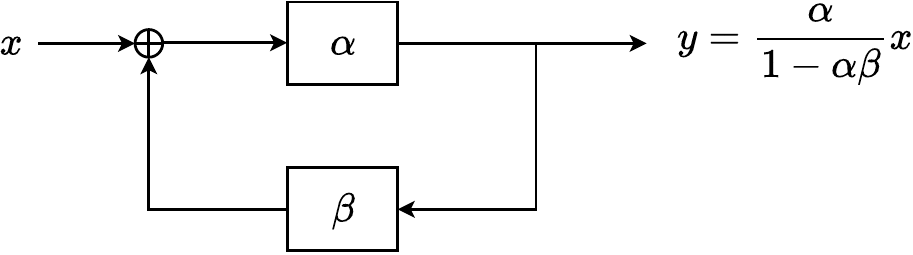}
		\caption{w/ self-interference}
		\label{subfig: w-self-interference}
	\end{subfigure}
	\caption{Repeater response with and without self-interference.}
	\label{fig: self-interference}
\end{figure}

Typically, when designing full-duplex devices, limiting the self-interference is an important design aspect to avoid self-oscillation and performance degradation.
However, we note that the meaning of self-interference is different in the context of full-duplex repeaters than in the context of
conventional full-duplex transceivers. In the full-duplex \emph{repeater} case, the repeater transmits \emph{the same signal} as it receives.
Hence, except the possibility of causing self-oscillation, the effect of self-interference in the repeater case is merely a change in effective gain.
On the other hand, none of our analyses excludes self-interference: in \eqref{eq: single repeater output}, the self-interference is included if $h_{nn}^\ttR \neq 0$, and all the results follow regardless.

To be specific, in Fig. \ref{fig: self-interference}, we illustrate the effective repeater response   without (a) and with (b) self-interference.
In either case, the repeater is a linear time-invariant system.
In (a), the repeater has a constant amplification gain $\alpha $, while in (b), the repeater also receives a self-interference signal with loopback gain $\beta$. 
From a system perspective, once self-oscillation is avoided by ensuring $\alpha < 1/\beta$, a repeater with self-interference can be treated as a repeater without self-interference but with an effective gain of $\alpha/(1-\alpha\beta)$.

\section{Numerical Analysis}
\label{sec: simulations}

The following default setup is used unless otherwise stated.
We consider both 6 GHz with 20 MHz bandwidth (referred to as FR1) and 30 GHz with 100 MHz bandwidth (referred to as FR2), representing two typical frequency bands in 5G \gls{nr}.
The simulation area is circular with radius 1000 meters for FR1 and 500 meters for FR2, where a \gls{bs} with $M=64$ antennas is located at the center and $K=20$ users are placed uniformly at random with 35 meters minimum distance to the \gls{bs}.
$N=40$ repeaters are placed approximately uniformly using hexagonal packing with 100 meters minimum distance to the \gls{bs}. 
We generate the pathloss according to the 3GPP models in \cite[Sec. 7.4.1]{TR38-901}: the direct and \gls{r2b} links follow the \gls{uma} model, and the \gls{u2r} and \gls{r2r} links follow the \gls{umi} model.
The height of the \gls{bs}, repeaters, and users are 25, 10, and 1.5 meters, respectively.
We assume that \gls{los} components always exist in \gls{r2b} links;\footnote{\label{ft: los}This is an optimistic assumption, and our aim is to show the potential performance gain. In practice, \gls{los} propagation may not always be guaranteed, but should be achieved with a high probability under careful deployment. 
We have also considered in Sec. \ref{sec: number of repeaters} the scenario where the repeaters are arbitrarily deployed so that \gls{los} links can exist with low probability.} for other links, the \gls{los} exists probabilistically according to \cite[Sec. 7.4.2]{TR38-901}.
The small-scale fading for non-\gls{los} components is modeled as Rayleigh fading.
The \gls{bs} has an antenna gain of 8 dBi, while the repeaters and users have isotropic antennas.
The maximum transmit power of users and repeaters are both set to $P_{\max} = P_{\max}^\ttR = 23$ dBm. 
The maximum amplification gain of repeaters is $A_{\max} = 90$ dB \cite{carvalho2024network}.
At the \gls{bs}, the noise spectral density is $-174$ dBm/Hz, and the noise figure is 9 dB.
The repeater has the same noise power as the \gls{bs}, i.e., $\varsigma_\ttR^2 / \varsigma_\ttB^2 = 1$.
The effect of self-interference is not considered in the simulations. However, as explained in Section \ref{sec: self-interference}, for each repeater, the effect of self-interference is only a change of the effective amplification gain from $\alpha$ to $\alpha/(1-\alpha\beta)$, where $\beta$ is the loopback channel amplitude.
Therefore, instead of optimizing $\alpha$, one may optimize the effective amplification gain $\alpha/(1-\alpha\beta)$ to account for self-interference.
We ignore the time-delays at repeaters in the simulations (which introduces phase shifts, as explained in Remark \ref{remark: time-delay}), while our analysis and optimization framework account for arbitrary time-delays.
In Algorithm \ref{alg: wmmse}, we use equal user weights so that the problem becomes sum-rate maximization.

Quantitative figures procured from our simulation scenario are:
(i) for FR1, the median of received \gls{snr} (pre-processing) at the \gls{bs} from cell-edge user and repeater are $-23.4$ and $-3.9$ dB (repeater noise ignored), respectively; 
and (ii) for FR2, the corresponding \glspl{snr} are $-32.6$ and $-15.1$ dB, respectively.
It can be seen that the cell-edge users experience very poor channel quality in both scenarios, making it difficult for them to perform reliable communication. The \gls{snr} further deteriorates in FR2 due to the increased pathloss and noise power resulting from the wider bandwidth.

\subsection{Repeater Placement}

To obtain some insights on repeater placement, we consider a scenario where a cell-edge user transmits with full power $P_{\max}$ and a repeater is moved along the line connecting the user and the \gls{bs}.
The \glspl{lsfc} are denoted by $\beta_\ttD$, $\beta_\ttU$, and $\beta_\ttB$ for the direct, \gls{u2r}, and \gls{r2b} links, respectively, which are calculated according to the aforementioned pathloss models. 
The effect of probabilistic \gls{los} is considered for direct and \gls{u2r} links by taking $\beta = p \beta^{\mathtt{LoS}} + (1-p) \beta^{\mathtt{NLoS}}$, where $p$ is the distance-dependent \gls{los} probability.
For an amplification gain $\alpha$, the received \gls{snr} at the \gls{bs} is ${P_{\max} (\beta_\ttD + \alpha^2 \beta_\ttU \beta_\ttB)} / (\varsigma_\ttB^2 + \alpha^2 \beta_\ttU \varsigma_\ttR^2 )$, where recall that $\varsigma_\ttB^2$ and $\varsigma_\ttR^2$ are the noise power at the \gls{bs} and repeater, respectively.
Compared to the direct-link \gls{snr}, ${P_{\max} \beta_\ttD} / {\varsigma_\ttB^2}$, the repeater can improve the \gls{snr} by a factor of
$
	\frac{1 + \alpha^2 \beta_\ttU \beta_\ttB / \beta_\ttD}{1 + \alpha^2 \beta_\ttU \varsigma_\ttR^2 / \varsigma_\ttB^2},
$
which monotonically increases with $\alpha^2$ when $\beta_\ttU / \varsigma_\ttR^2 \geq \beta_\ttD / \varsigma_\ttB^2$, and decreases otherwise.
This implies that, to maximize the \gls{snr}, the repeater amplification gain should be selected as
\begin{equation}
	\label{eq: amplification gain for maximized SNR}
	\alpha = 
	\left\{
	\begin{array}{ll}
		\displaystyle\min\left\{ A_{\max}, \sqrt{\frac{P_{\max}^\ttR}{P_{\max} \beta_\ttU + \varsigma_\ttR^2}} \right\}, & \displaystyle\frac{\beta_\ttU}{\varsigma_\ttR^2} \geq \frac{\beta_\ttD}{\varsigma_\ttB^2} \\
		0, & \mbox{otherwise}.
	\end{array}
	\right.
\end{equation}

\begin{figure}
	\centering
	\begin{subfigure}[b]{0.45\textwidth}
		\centering
		\includegraphics[width=\textwidth]{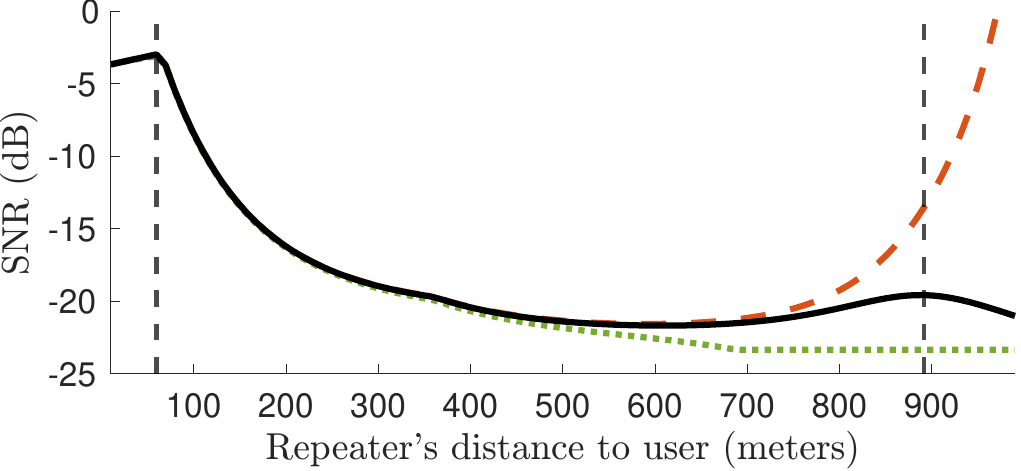}
		\caption{Uplink}
		\label{subfig: placement uplink}
	\end{subfigure}
	\hfill
	\begin{subfigure}[b]{0.45\textwidth}
		\centering
		\includegraphics[width=\textwidth]{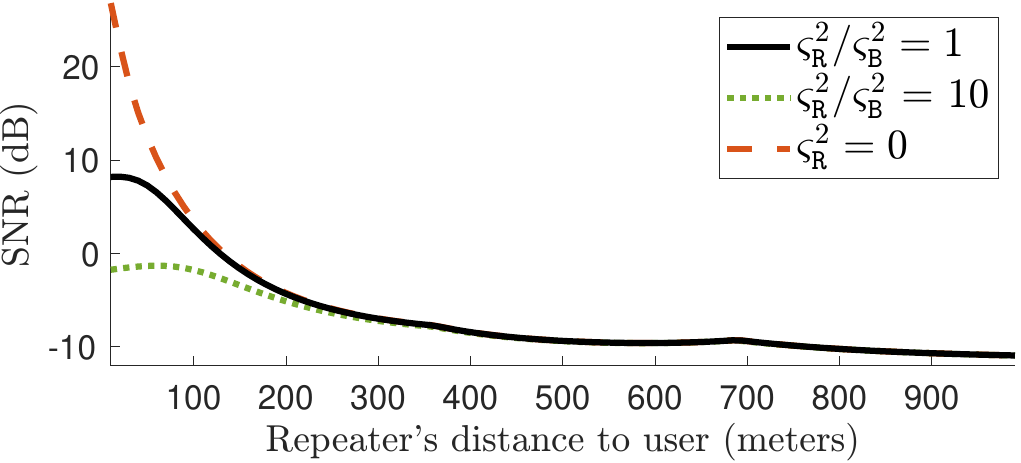}
		\caption{Downlink}
		\label{subfig: placement downlink}
	\end{subfigure}
	\caption{SNR versus repeater location.}
	\label{fig: repeater placement}
\end{figure}

In Fig. \ref{subfig: placement uplink}, we plot the uplink \gls{snr} at the \gls{bs} for FR1 when the repeater is placed at different locations along the line connecting the user and the \gls{bs} under different repeater noise levels.
When $\varsigma_\ttR^2 / \varsigma_\ttB^2 = 1$, we observe two transition points---marked by vertical dashed lines---where the \gls{snr} trend changes.
Before the first transition point, the repeater is too close to the user and the amplification gain is limited by the repeater power constraint.
Between the two transition points, the concatenated user-repeater-\gls{bs} link strength is restricted by the ``double pathloss'', i.e., both user-to-repeater and repeater-to-\gls{bs} links are weak, making the product, $\beta_\ttU \beta_\ttB$, extremely small.
Beyond the second transition point, the user-to-repeater \gls{snr}, $\beta_\ttU / \varsigma_\ttR^2$, becomes very low, making it counter-productive to move the repeater further to the \gls{bs}.
Contrarily, in the other two cases: (i) when there is no noise at the repeater ($\varsigma_\ttR^2 = 0$), the \gls{snr} keeps increasing when moving the repeater closer to the \gls{bs}; and (ii) when the repeater noise level is high ($\varsigma_\ttR^2 / \varsigma_\ttB^2 = 10$), the repeater has to be turned off if it is too far away from the user, since the signal at the repeater is already too noisy.

The results for downlink are shown in Fig. \ref{subfig: placement downlink}, where the maximum transmit power of the \gls{bs} is set to 35 dBm. (The mathematical analysis is similar to uplink, and is therefore omitted.)
We observe that, also in the downlink, placing the repeater close to the user is more beneficial.
This is because the \gls{bs} has much higher transmit power, and a repeater close to the \gls{bs} can provide little additional gain.
Since we consider that the repeaters always have \gls{los} with the \gls{bs}, it can receive a much stronger signal than the user even at the cell edge.

The above results suggest that the repeaters should be placed such that they receive strong signals from users at the cell edge or coverage holes.
When the pathloss information is unknown or time-varying due to mobility, a uniform placement of repeaters is a practical choice, which increases the likelihood of a user having a strong channel to a nearby repeater.
Interestingly, such placement in turn maximizes the inter-repeater spacing, enhancing the stability of the repeater swarm.

\subsection{Convergence of the Optimization Algorithm}

\begin{figure}
	\centering
	\begin{subfigure}[b]{0.45\textwidth}
		\centering
		\includegraphics[width=\textwidth]{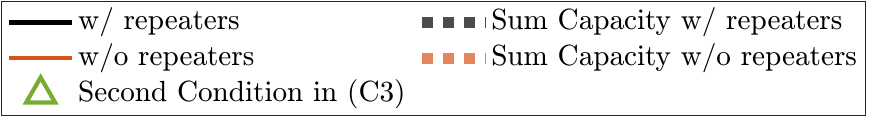}
	\end{subfigure}
	\hfill
	\begin{subfigure}[b]{0.465\textwidth}
		\centering
		\includegraphics[width=\textwidth]{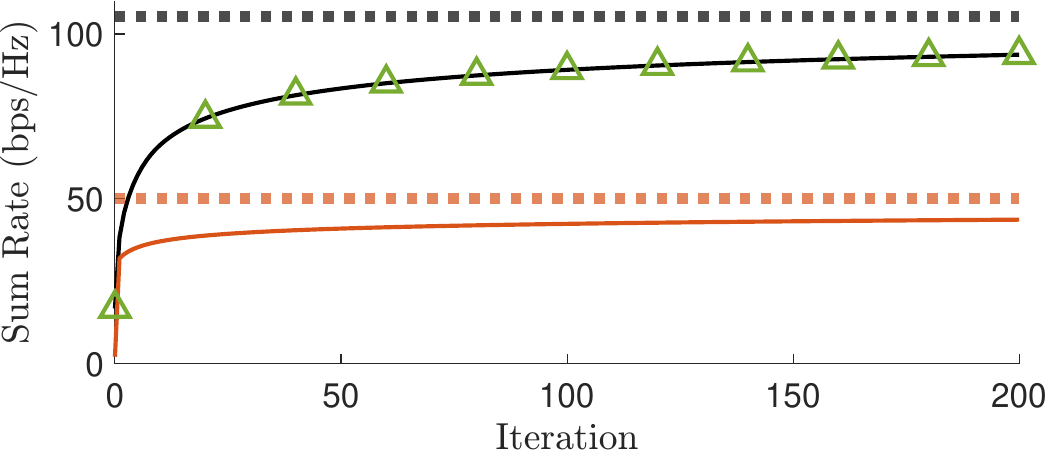}
		\caption{FR1 (6GHz, 20 MHz bandwidth)}
		\label{subfig: convergence FR1}
	\end{subfigure}
	\hfill
	\begin{subfigure}[b]{0.45\textwidth}
		\centering
		\includegraphics[width=\textwidth]{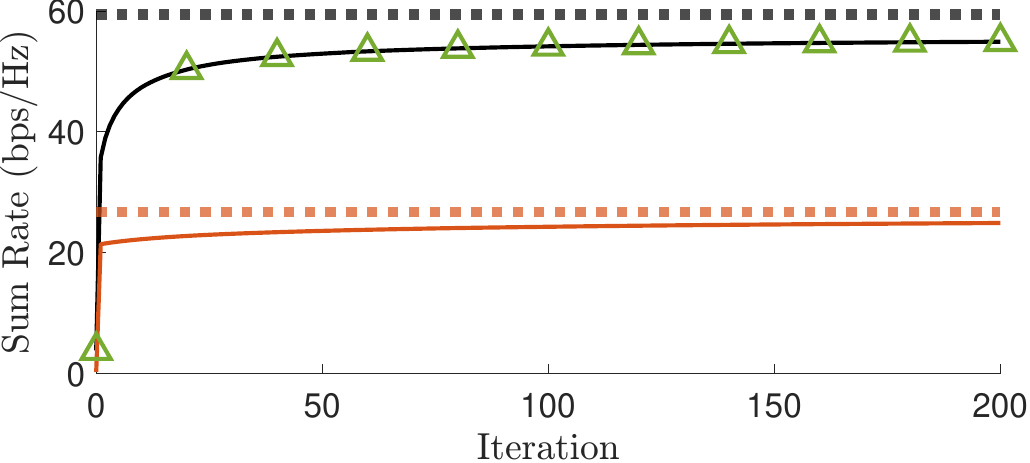}
		\caption{FR2 (30 GHz, 100 MHz bandwidth)}
		\label{subfig: convergence FR2}
	\end{subfigure}
	\caption{Convergence of the proposed Algorithm \ref{alg: wmmse}.}
	\label{fig: convergence}
\end{figure}

Next, to demonstrate the convergence of Algorithm~\ref{alg: wmmse}, we plot the sum rate achieved per iteration in Fig. \ref{fig: convergence}, with or without repeaters~(where Line \ref{alg-line: repeater gain} in Algorithm \ref{alg: wmmse} is skipped in  the latter case).
We choose the first condition in the repeater stability constraint \eqref{C3} and set $\eta = 0.9$. 
(Notice that the result obtained using the second condition in \eqref{C3} instead---plotted with triangle markers---is almost identical.)
The repeater power constraint \eqref{C7} is included when updating the repeater gains.
Convergence of the algorithm can be readily observed: the sum rate consistently increases in the beginning and saturates (with a negligible diminishing return) with the progress of iterations. 
However, more iterations are needed for the algorithm to converge in FR1 compared to FR2.
This is because the pathloss is much lower in FR1, and many more communication links have non-negligible effects. 
To balance complexity and performance, we henceforth set the maximum number of iterations to $I_{\max} = 50$, and enable early stopping if the sum rate improvement is less than $\epsilon = 10^{-3}$ bps/Hz between consecutive iterations for the simulation results.
As a reference, we also plot the sum capacity in \eqref{eq: sum capacity} using the repeater gains obtained after the algorithm converges. 
Note that the repeater gains are optimized for sum rate, which do not necessarily maximize the sum capacity---nevertheless, this should work reasonably well, as we observe that the achieved sum rate is quite close to the sum capacity.

\subsection{Number of Repeaters}
\label{sec: number of repeaters}

\begin{figure}
	\centering
	\begin{subfigure}[b]{0.3\textwidth}
		\centering
		\includegraphics[width=\textwidth]{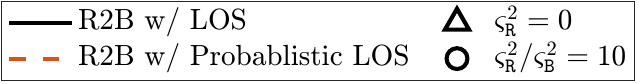}
	\end{subfigure}
	\hfill
	\begin{subfigure}[b]{0.45\textwidth}
		\centering
		\includegraphics[width=\textwidth]{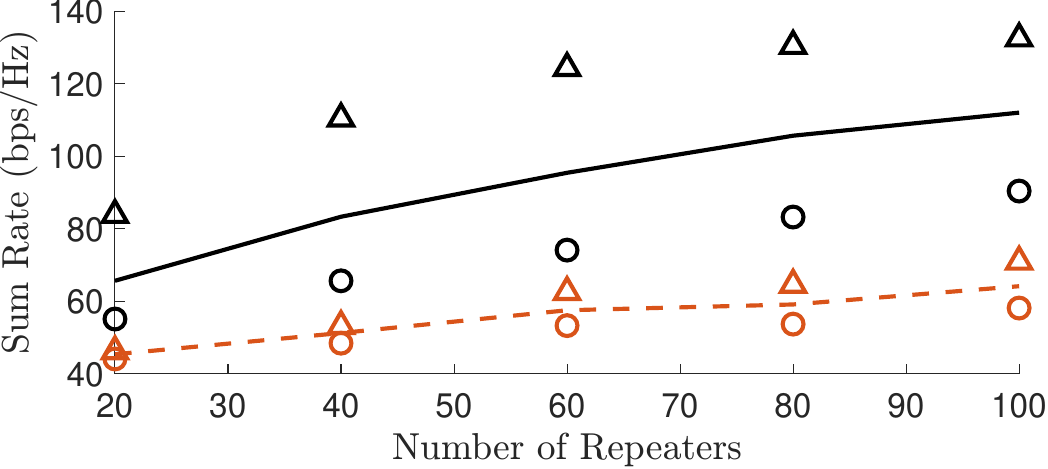}
		\caption{FR1 (6 GHz, 20 MHz bandwidth)}
		\label{subfig: number of repeaters uplink}
	\end{subfigure}
	\hfill
	\begin{subfigure}[b]{0.45\textwidth}
		\centering
		\includegraphics[width=\textwidth]{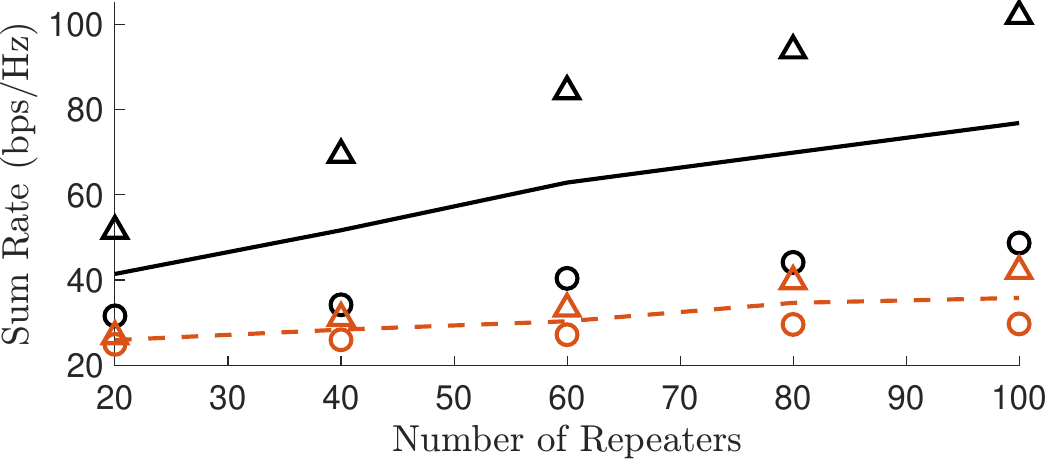}
		\caption{FR2 (30 GHz, 100 MHz bandwidth)}
		\label{subfig: number of repeaters downlink}
	\end{subfigure}
	\caption{Sum rate versus number of repeaters.}
	\label{fig: number of repeaters}
\end{figure}

\begin{figure}
	\centering
	\begin{subfigure}[b]{0.45\textwidth}
		\centering
		\includegraphics[width=\textwidth]{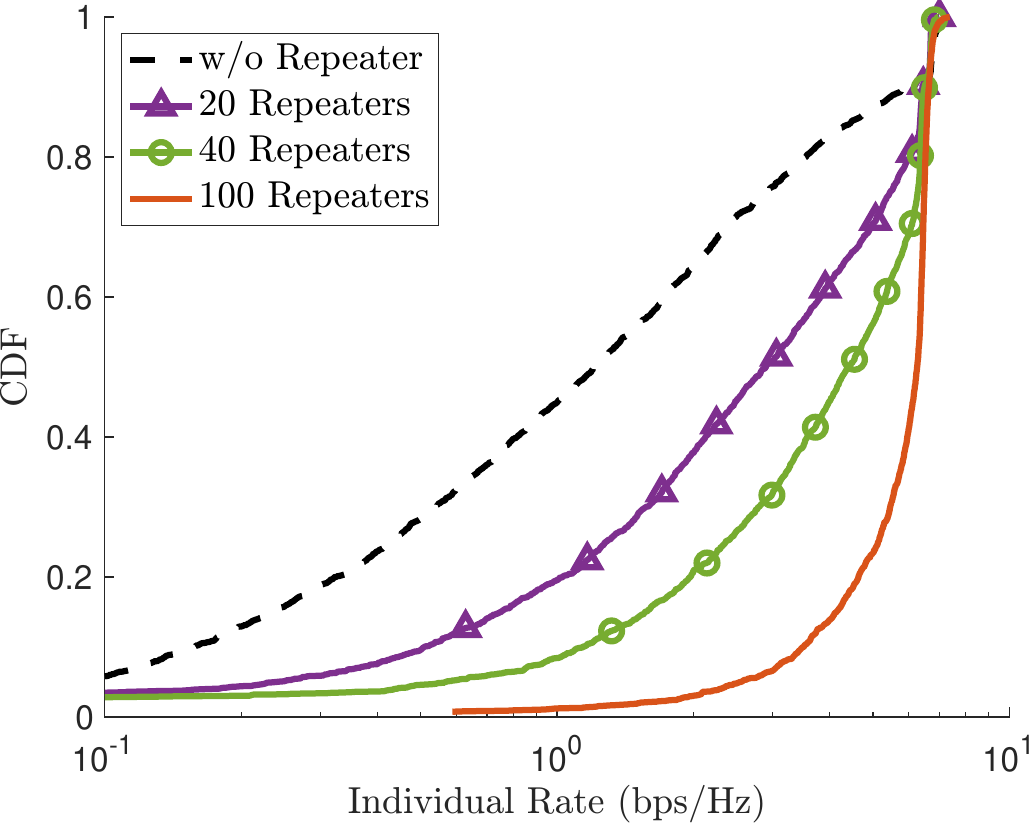}
		\caption{FR1 (6GHz, 20 MHz bandwidth)}
		\label{subfig: individual rate uplink}
	\end{subfigure}
	\hfill
	\begin{subfigure}[b]{0.45\textwidth}
		\centering
		\includegraphics[width=\textwidth]{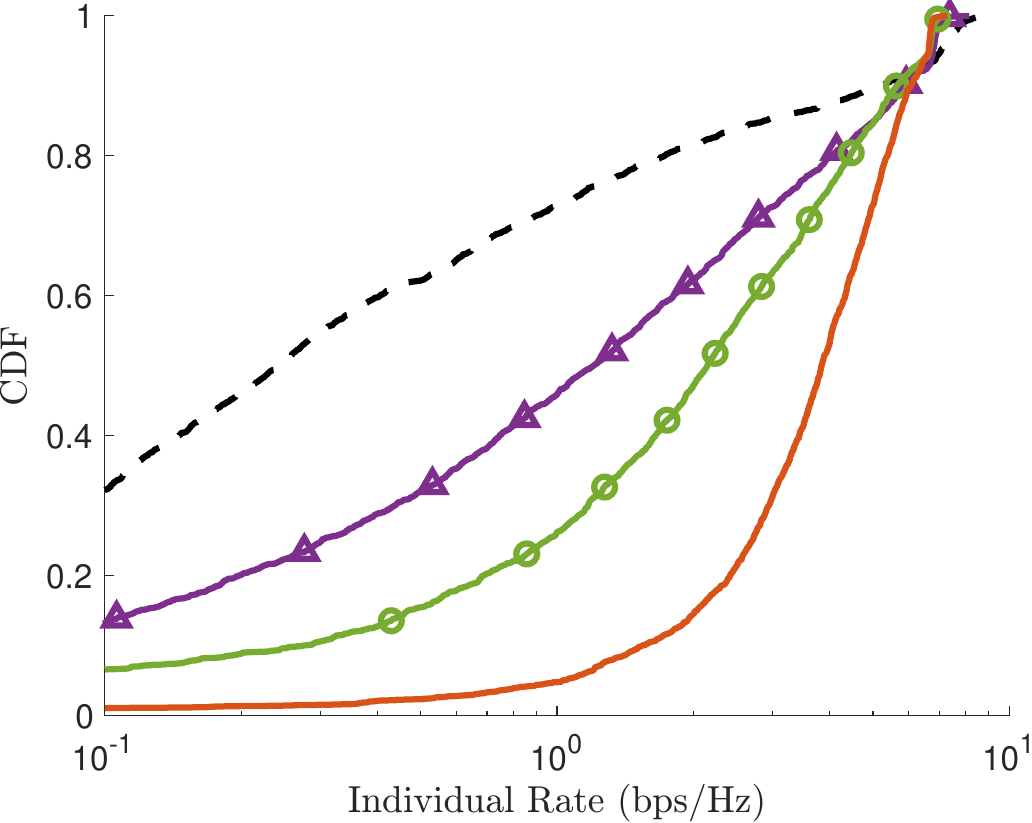}
		\caption{FR2 (30 GHz, 100 MHz bandwidth)}
		\label{subfig: individual rate downlink}
	\end{subfigure}
	\caption{Individual rate distribution.}
	\label{fig: individual rate}
\end{figure}

The achieved sum rate with different number of repeaters is plotted in Fig. \ref{fig: number of repeaters}.
Particularly, we compare the results with the case where \gls{los} exists only probabilistically in the repeater-to-\gls{bs} links (notice that the \gls{los} probability is already less than 5\% when the repeater is 400 meters away from the \gls{bs}).
As observed, the performance gains from repeaters are largely negated when the \gls{los} cannot be guaranteed.
It is necessary to ensure that the repeaters always have strong channels to the \gls{bs} through careful placement.
The noise power at the repeater is also a critical factor, as the performance deteriorates significantly when the repeater noise level is very high.

In Fig. \ref{fig: individual rate}, we plot the \gls{cdf} of individual user rates with different numbers of repeaters.
As observed, a denser repeater deployment provides a more uniform rate distribution among users, as even cell-edge users can achieve an improved rate with one or more repeaters in the vicinity.
We also observe an ``implicit scheduling'' effect (a small portion, $\approx 2.5\%$, of users have zero transmit power) when the number of repeaters is not so large.
This is because when two users are equally close to an isotropic-antenna repeater, they cause strong interference to each other if both transmit (please recall the example in Sec.~\ref{sec: example}).
Since the chosen optimization criterion is sum-rate maximization, one user may be silenced for the other to transmit more effectively.
However, this should not be a significant issue due to the following reasons:
(i) compared to the case without repeaters, the number of users with very low rates (e.g., less than 0.1 bps/Hz) is still less; 
(ii) the problem can be avoided by user scheduling, which is always implemented in the systems and will not require much additional overhead due to the small number of affected users; 
(iii) although short-term fairness may be compromised, sum-rate maximization still provides long-term fairness, as the user location will change over time;
and (iv) users with high quality-of-service requirements can be prioritized by assigning them higher weights in the optimization.
Alternatively, one could consider different optimization criteria (e.g., max-min fairness) or use more advanced multi-antenna repeaters for spatial separability, which is beyond the scope of this paper.

\subsection{Impact of Stability Margin $\eta$}
\label{sec: stability margin}

\begin{figure}
	\centering
	\includegraphics[width=0.45\textwidth]{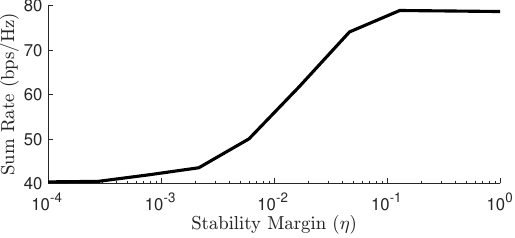}
	\caption{Impact of the stability margin $\eta$.}
	\label{fig: stability margin}
\end{figure}

In the stability condition \eqref{C3}, we introduced a stability margin $\eta$ to provide robust stability under channel uncertainties.
It is clear that a smaller $\eta$ leads to a more conservative stability condition, which may limit the amplification gain of repeaters and, consequently, the achievable sum rate.
Ideally, $\eta$ should be chosen such that the impact of inter-repeater interaction can be sufficiently mitigated, while not being too conservative to avoid unnecessary performance loss.
Unfortunately, it is not straightforward to accurately analyze the impact of $\eta$ on the sum rate.
Instead, in Fig. \ref{fig: stability margin}, we numerically investigate the impact of $\eta$ on the sum rate in FR1.
We observe that the result is not sensitive to the choice of $\eta$ as long as it is not too small (e.g., $\eta < 0.1$).
However, we have to note that the results depend on the chosen simulation setup, and may vary in different scenarios.

\section{Concluding Remarks}

Deploying swarms of low-cost, low-complexity, and low-power repeaters is a promising solution to improve coverage and channel rank in cellular networks.
For example, in our tested scenario, deploying 40 single-antenna repeaters---each with the same power budget as user devices---within a cell of 1000 meters radius can nearly double the sum rate compared to the case without repeaters.

However, to achieve the full potential of the envisioned repeater swarm-assisted cellular system, careful system design and optimization are necessary. 
First, repeaters inject additional noise and interference.
Second, interaction instability caused by the positive feedback between repeaters must be avoided---this can be guaranteed by verifying the sufficient conditions for stability that we derived; these conditions depend only on the inter-repeater channel amplitudes.
Third, the repeaters should be placed close to the users that require assistance, for example, at the cell edge and coverage holes, to combat the ``double pathloss'' effect and the injected noise.
Preferably, the repeaters should have strong \gls{los} channels to the \gls{bs} through careful placement.

Considering multi-cell scenarios with imperfect channel state information and pilot contamination can be of   interest for future study.
Other potential directions include generalization to time-varying systems, the optimization of repeater time-delays, more effective signaling schemes, a deeper analysis of self-interference, and the use of multi-antenna repeaters to enhance spatial separability.

\appendix
\section*{Proof of Theorem \ref{th: stability}}
\label{app: stability proof}

Conditions for the existence of a unique and causal impulse response are provided in \cite[Th. 28.2]{doetsch2012introduction}, and the condition for the bounded-energy stability is given in \cite[Th. 4.3]{zhou1998essentials}.
Our proof idea is to show that---under the  conditions imposed in Theorem \ref{th: stability}---the requirements of both theorems are satisfied for our particular system.

Instead of $\bG(s)$, we consider the transfer function matrix
\begin{align}
	\bZ(s) 
	=& \bG(s) - \bD_{\ba(s)} \notag \\ 
	=& \left(\bI - \bD_{\ba(s)} \bH^\ttR(s) \right)^{-1} \bD_{\ba(s)}\bH^\ttR(s)\bD_{\ba(s)}. \label{eq: Z}
\end{align}
Since the system $\bD_{\ba(s)}$ is causal and stable (each diagonal element represents a finite amplification with delay), $\bG(s)$ is causal and stable if and only if $\bZ(s)$ is causal and stable.

To proceed, we prove the following lemma.

\begin{lemma}
	\label{lemma}
	Under Assumptions \ref{as: analyticity} and \ref{as: channel amplitude gain}, if the image of $\det(\bI - \bD_{\ba(j\omega)} \bH^\ttR(j\omega))$ does not encircle the origin, then
	\begin{enumerate}[label=(\alph*),font=\normalshape]
		\item 
		\label{assertion: analyticity}
		$\bZ(s)$ is analytic in $\C_{+}$ 
		\item 
		\label{assertion: convergence}
		All elements of $\bZ(s)$ converges to 0 as $|s|\rightarrow \infty$ in $\C_+$ 
		\item All elements of $\bZ(s)$ are absolutely integrable on all vertical lines in $\C_+$. \label{assertion: integrability}
	\end{enumerate}
\end{lemma}

\begin{IEEEproof}
	\ref{assertion: analyticity}
	Since both $\bH^\ttR(s)$ and $\bD_{\ba(s)}$ are analytic in $\C_+$ and the determinant of a matrix involves only sums and products (both of which preserve analyticity) of its elements, $\det(\bI_N - \bD_{\ba(j\omega)} \bH^\ttR(j\omega))$ is analytic in $\C_+$ and, therefore, has no pole in $\C_+$.
	Consider the Nyquist contour, consisting of a path traveling up the $j\omega$ axis, from $0-j\infty$ to $0+j\infty$, along with a semicircular arc in $\C_+$ of infinitely large radius that starts at $0+j\infty$ and travels clockwise to $0-j\infty$, enclosing the entire $\C_+$. 
	According to Cauchy's  argument principle \cite[pp. 230]{brown2009complex},  the difference between the number of zeros and poles of $\det(\bI_N - \bD_{\ba(s)} \bH^\ttR(s) )$ within $\C_+$ equals the number of clockwise encirclements of the origin traced by its image as $s$ traverses the Nyquist contour in a clockwise direction.  
	Since $\det(\bI_N - \bD_{\ba(s)} \bH^\ttR(s) )$ has no poles in $\C_+$, the number of encirclements of the origin equals the number of zeros.  
	Furthermore, under Assumption \ref{as: channel amplitude gain}, $|h_{nn'}(s)| \rightarrow 0$ as $|s|\rightarrow 0$ in $\C_+$, meaning that on the semicircular part of the contour, the image of $\det(\bI_N - \bD_{\ba(s)} \bH^\ttR(s) )$ collapses to a single point at $1 + j0$.
	Thus, if the image of $\det(\bI_N - \bD_{\ba(j\omega)}\bH^\ttR(j\omega))$ does not encircle the origin, it follows that $\det(\bI_N - \bD_{\ba(s)} \bH^\ttR(s) )$ has no zero in $\C_+$, and we have
	\begin{equation}
		\left(\bI_N - \bD_{\ba(s)} \bH^\ttR(s) \right)^{-1} = \frac{\adj\left(\bI_N - \bD_{\ba(s)} \bH^\ttR(s) \right)}{\det\left(\bI_N - \bD_{\ba(s)} \bH^\ttR(s) \right)}.
	\end{equation}
	Since all elements in the adjugate matrix are determinants of some sub-matrices (possibly with sign changes), the adjugate matrix $\adj(\bI_N - \bD_{\ba(s)} \bH^\ttR(s) )$ is also analytic in $\C_+$.
	As division preserves analyticity when the denominator is nonzero, we have that $(\bI_N - \bD_{\ba(s)} \bH^\ttR(s) )^{-1}$ is analytic in $\C_+$.
	Then, from \eqref{eq: Z}, we have that $\bZ(s)$ is analytic in $\C_+$.

	\ref{assertion: convergence}-\ref{assertion: integrability}
	For an arbitrary matrix $\bX$, we denote by $\sigma_n(\bX)$ the $n$th singular value, by $\sigma_{\min}(\bX)$ the smallest singular value, by $\matnorm{\bX}_2$ the spectral norm (i.e., the largest singular value), and by $\norm{\bX}_\infty$ the maximum magnitude of its elements (i.e., the $\ell_\infty$ vector norm applied after vectorizing $\bX$).
	For $|s| > \delta$ in $\C_+$, all singular values of $\bI - \bD_{\ba(s)} \bH^\ttR(s)$ satisfy
	\begin{align}
		\abs{\sigma_n\left(\bI - \bD_{\ba(s)} \bH^\ttR(s) \right) - 1}
		\leq&~ \matnorm{\bD_{\ba(s)} \bH^\ttR(s)}_2 \notag \\
		\leq&~ N \norm{\bD_{\ba(s)} \bH^\ttR(s)}_{\infty} \notag \\
		\leq&~ \frac{NCA}{|s|^{1+\varepsilon}}, \label{eq: diff in singular values}
	\end{align}
	where $A \defeq \max_{n} \alpha_n$ is the maximum amplification gain among all repeaters.
	In \eqref{eq: diff in singular values}, the first inequality follows from \cite[Cor. 7.3.5]{horn2012matrix}, the second inequality from $\matnorm{\bX}_2 \leq N \norm{\bX}_\infty$ \cite[pp. 365]{horn2012matrix}, and the last inequality from Assumption \ref{as: channel amplitude gain}.
	From \eqref{eq: diff in singular values}, we have 
	\begin{equation}
	\label{eq: lower bound on singular values}
		\sigma_{\min}\left(\bI - \bD_{\ba(s)} \bH^\ttR(s)\right) \geq 1 - \frac{NCA}{|s|^{1+\varepsilon}}.
	\end{equation}
	Thus, for $|s| \geq  \delta' \defeq \max\{\delta, (2NCA)^{\frac{1}{1+\varepsilon}}\}$ in $\C_+$,
	\begin{align}
		\norm{\left(\bI - \bD_{\ba(s)} \bH^\ttR(s)\right)^{-1}}_{\infty}
		&\leq \matnorm{\left(\bI - \bD_{\ba(s)} \bH^\ttR(s)\right)^{-1}}_2 \notag \\
		&= \frac{1}{\sigma_{\min}\left(\bI - \bD_{\ba(s)} \bH^\ttR(s)\right)} \notag \\
		&\leq \frac{1}{\displaystyle 1 - \frac{NCA}{|s|^{1+\varepsilon}}} \notag \\
		&\leq 1 + \frac{2NCA}{|s|^{1+\varepsilon}}, 	\label{eq: upper bound on max norm}
	\end{align}
	where the first inequality follows from $\norm{\cdot}_{\infty} \leq \matnorm{\cdot}_2$ \cite[pp. 365]{horn2012matrix}, the second inequality from \eqref{eq: lower bound on singular values}, and the last inequality is obtained by applying the inequality $\frac{1}{1-x} = 1 + \frac{x}{1-x} \leq 1 + 2x$ with $x = \frac{NCA}{|s|^{1+\varepsilon}} \in (0,\frac{1}{2}]$ which holds if $|s| \geq (2NCA)^{\frac{1}{1+\varepsilon}}$.
	Now, for $|s| \geq \delta'$ in $\C_+$, we have
	\begin{align}
		&\norm{\bZ(s)}_{\infty}
		= \norm{\left(\bI - \bD_{\ba(s)} \bH^\ttR(s) \right)^{-1} \bD_{\ba(s)}\bH^\ttR(s)\bD_{\ba(s)}}_{\infty} \nonumber \\
		&\leq N \norm{\left(\bI - \bD_{\ba(s)} \bH^\ttR(s) \right)^{-1}}_{\infty} \norm{\bD_{\ba(s)}\bH^\ttR(s)\bD_{\ba(s)} }_{\infty} \nonumber \\
		&\leq N \left(1 + \frac{2NCA}{|s|^{1+\varepsilon}} \right)  \frac{CA^2}{|s|^{1+\varepsilon}} \nonumber \\
		&= \frac{NCA^2}{|s|^{1+\varepsilon}} + \frac{2N^2C^2A^3}{|s|^{2+2\varepsilon}},
		\label{eq: upper bound on max norm of Z}
	\end{align}
	where the first inequality is due to the fact that $N\norm{\cdot}_{\infty}$ is a matrix norm and satisfies the submultiplicativity property, and the second inequality can be obtained by applying \eqref{eq: upper bound on max norm} and Assumption \ref{as: channel amplitude gain}.
	From \eqref{eq: upper bound on max norm of Z}, it follows that $\norm{\bZ(s)}_{\infty}$ converges to 0 as $|s|\rightarrow \infty$ in $\C_+$ and 
	$$
		\int_{-\infty}^{-\delta'} \norm{\bZ(\sigma + j\omega)}_{\infty} d\omega + \int_{\delta'}^\infty \norm{\bZ(\sigma + j\omega)}_{\infty} d\omega < \infty.
	$$
	From \ref{assertion: analyticity}, we have that $\bZ(s)$ is analytic in $\C_+$; therefore, 
	$$
	\int_{-\delta'}^{\delta'} \norm{\bZ(\sigma + j\omega)}_{\infty} d\omega < \infty.
	$$
	Combining the above results, we conclude that all elements of $\bZ(s)$ are absolutely integrable on all vertical lines in $\C_+$.

\end{IEEEproof}

When all the assertions in Lemma \ref{lemma} hold, it follows from \cite[Th. 28.2]{doetsch2012introduction} that $\bZ(s)$ represents the Laplace transform of the impulse response
\begin{equation}
	\mat{\sfZ}(t) \defeq \frac{1}{j2\pi} \int_{\sigma-j\infty}^{\sigma+j\infty} \bZ(s) e^{s t}~ \mathrm{d} s,
\end{equation}
which is unique, i.e., the integral is independent of the choice of $\sigma$ when $\sigma \geq 0$, and causal, i.e., $\mat{\sfZ}(t) = \bzero$ for $t<0$.

Furthermore, $\bZ(s)$ has a finite $\cH_\infty$ norm, i.e.,
\begin{equation}
	\|\bZ\|_{\cH_\infty} \defeq \sup_{\Re\{s\}>0} \matnorm{\bZ(s)}_2 < \infty,
\end{equation} 
since $\bZ(s)$ is analytic in $\C_+$ and $\norm{\bZ(s)}_{\infty}$ converges to zero as $|s| \rightarrow \infty$ in $\C_+$.
According to \cite[Th. 4.3]{zhou1998essentials}, $\mat{\sfZ}(t)$ is bounded energy stable when $\|\bZ\|_{\cH_\infty}$ is finite.
As previously mentioned, $\bG(s)$ shares the same stability properties as $\bZ(s)$, which completes the proof.

\bibliographystyle{IEEEtran}
\bibliography{ref}

\end{document}